\documentclass{article}

\usepackage{fieldnotes}

\usepackage[utf8]{inputenc}
\usepackage[T1]{fontenc}
\usepackage[hidelinks]{hyperref}
\usepackage{url}
\usepackage{amsmath,amsfonts}
\usepackage{microtype}
\usepackage{graphicx}

\usepackage{booktabs,tabularx,longtable,array,ragged2e}
\newcolumntype{L}[1]{>{\RaggedRight\arraybackslash}p{#1}}
\newcolumntype{Y}{>{\RaggedRight\arraybackslash}X}

\usepackage{tikz}
\usetikzlibrary{arrows.meta,positioning}

\usepackage[backend=biber,style=numeric]{biblatex}

\preprintlabel{Preprint}

\preprintmeta{March 2026}
\runningtitle{Topology as a Language for Emergent Organization in Complex Systems}

\title{Topology as a Language for Emergent Organization in Complex Systems:
Multiscale Structure, Higher-Order Interactions, and Early Warning Signals}

\author{
  \textbf{Mark M. Bailey} \\
  AI, Cyber, Influence, and Data Science Department \\
  Biological and Computational Intelligence Center \\
  National Intelligence University \\
  Bethesda, MD, USA \\
  \texttt{mark.m.bailey@odni.gov}
}

\begin{document}
\maketitle

\begin{abstract}
Complex systems are difficult to study not only because they are nonlinear, multiscale, and often nonstationary, but because their scientifically relevant organization is often invisible at the level of individual components, pairwise interactions, or low-order summary statistics. This review argues that topology has become valuable in complex-systems science because it provides a mathematical language for representing emergent organization when relevant structure is distributed, relational, and robust across scale. We synthesize work on persistent homology, Mapper, simplicial complexes, hypergraphs, and related operators, while distinguishing invariant-based topological methods from broader topology-inspired representations. We show how persistence formalizes multiscale stability, how higher-order models preserve collective interactions erased by pairwise graphs, and how topological approaches complement rather than replace statistics, graph theory, and geometry. We review applications in nonlinear dynamics, neuroscience, finance, ecology, materials science, and anomaly detection, emphasizing a common logic: topology turns reorganizing structure into measurable signals for regime shifts, state transitions, and early warning. Across domains, these methods are most effective when the scientific target is organizational rather than scalar, when threshold ambiguity is intrinsic to the problem, and when topology functions as a structural diagnostic or feature extractor within a broader analytic pipeline. We conclude by identifying key limitations, including representation dependence, inferential challenges, interpretability, computational scaling, and the narrowness of one-parameter workflows, and by outlining a research agenda linking topology more closely to dynamics, causality, streaming decision support, topology-aware AI, and socio-technical resilience.
\end{abstract}



\section{Introduction: The Representation Problem in Complex Systems}
\label{sec:introduction}

Complex systems science is united less by a single formalism than by a recurring family of questions: how to represent systems composed of many interacting parts, how to relate local interaction rules to macroscopic order, and how to describe structures that are distributed across scales rather than concentrated in a few variables. Simon's classic account of hierarchy and near-decomposability, Anderson's argument that ``more is different,'' and Ladyman, Lambert, and Wiesner's later attempt to clarify what qualifies as a complex system all converge on a shared point: the scientifically interesting features of complex systems are rarely exhausted by a listing of components or by a specification of local rules alone \cite{Simon1962a,Anderson1972a,Ladyman2013a}. The problem is therefore not only explanatory but representational. Before one can ask why a complex system adapts, fails, or undergoes abrupt transition, one must first decide what mathematical object is supposed to stand in for the system \cite{Torres2021a}.

That decision is consequential because every representation foregrounds some forms of structure and suppresses others. Statistical descriptions excel at summarizing distributions, moments, and correlations; dynamical-systems approaches clarify trajectories, attractors, bifurcations, and stability; network models encode interaction patterns that would otherwise remain hidden. None of these traditions is dispensable, and each has generated major advances in the study of ecosystems, brains, markets, infrastructures, and social systems \cite{Scheffer2009a,Torres2021a}. The difficulty is not that the dominant formalisms are wrong, but that they are incomplete. In many complex systems, what matters most is not simply the magnitude of a variable or the presence of an edge, but the organization of relations across scales: recurrent pathways, collective constraints, mesoscale coordination, and higher-order interaction patterns that do not reduce cleanly to low-order summaries.

Recent work on higher-order networks makes this limitation especially vivid. Pairwise graphs have been extraordinarily productive, but an expanding literature now shows that many systems---from social communication to biochemical reactions and neural coordination---involve interactions among three or more units that cannot be faithfully reduced to dyads without changing the effective dynamics of contagion, synchronization, diffusion, or collective decision-making \cite{Battiston2020a,Bick2023a,Majhi2022a,Boccaletti2023a,Zhang2023a}. Representation is therefore not a neutral preprocessing step. As Torres and colleagues emphasize, the choice of representation shapes which observables become available, which hypotheses can be posed, and even which kinds of explanation appear plausible \cite{Torres2021a}. If the phenomenon of interest is fundamentally organizational rather than merely scalar, then the representation must be able to preserve organization as a first-class scientific object.

This is the point at which topology becomes attractive. At its broadest, topology concerns qualitative features of spaces and relations that persist under continuous deformation. In data analysis, that perspective has matured into a family of tools---most prominently persistent homology, but also Mapper, Reeb-graph-like constructions, and simplicial representations---designed to extract robust structural information from noisy, high-dimensional, or relational data \cite{Carlsson2009a,Edelsbrunner2008a,Wasserman2018a}. Within the complex-systems literature itself, Merelli and colleagues explicitly argued that persistence-based summaries can serve as descriptors of system organization rather than as generic machine-learning features \cite{Merelli2015a}. The central insight is simple but powerful: a system may change in metric detail while preserving a deeper pattern of connectedness, recurrence, exclusion, or cavity structure. When the scientific question concerns the persistence of that deeper pattern, topological methods become more than simply descriptive. They provide observables tailored to the qualitative organization of the data.

Persistent homology is especially important in this regard because it formalizes the idea that structure should be interrogated across scale rather than read off at a single threshold. Instead of treating a dataset, weighted network, or reconstructed state space as fixed once and for all, persistent homology tracks how connected components, loops, and higher-dimensional cavities appear and disappear along a filtration \cite{Edelsbrunner2008a,Carlsson2009a}. Cohen-Steiner, Edelsbrunner, and Harer showed that persistence diagrams enjoy stability properties under perturbation, helping explain why persistent features can be interpreted as robust signatures rather than arbitrary artifacts of sampling or noise \cite{CohenSteiner2007a}. Otter and colleagues later systematized the computational pipeline, while Bubenik showed how persistence landscapes can embed topological summaries in statistical workflows \cite{Otter2017a,Bubenik2015a}. For complex systems, where noise, incomplete observation, and heterogeneous scales are the rule rather than the exception, this combination of qualitative abstraction and quantitative robustness is especially appealing.

The significance of topology in the present review, however, is not exhausted by persistent homology narrowly construed. The word ``topological'' is used here in a broad but disciplined sense. Persistent invariants and their derived summaries form the conceptual core, but simplicial complexes and related higher-order representations are also included insofar as they preserve joint interaction structure that ordinary graphs erase \cite{Battiston2020a,Bick2023a}. The objects on which topological methods operate need not be ordinary point clouds; they can be built from time-delay embeddings, weighted correlation networks, adjacency relations, clique expansions, or explicitly higher-order interaction data \cite{Horak2009a,Stolz2017a,Gidea2018a}. In this broader sense, topological methods are useful not only because they summarize ``shape'' in an abstract way, but because they permit the analyst to represent a system at the level where collective constraints and joint interactions actually occur.

That point leads directly to emergence. Anderson's well-known claim that ``more is different'' was not simply a slogan against reductionism; it was an argument that higher levels of organization possess regularities that demand their own descriptive vocabulary \cite{Anderson1972a}. Simon's discussion of hierarchical organization made a parallel point by showing that complex systems are often intelligible only when their nested structure is made explicit \cite{Simon1962a}. Crutchfield later argued that the detection of structure depends crucially on the model class available to the observer, while Rupe and Crutchfield more recently emphasized that the physics of self-organization still suffers from a chronic failure to define and compare structure rigorously \cite{Crutchfield1994a,Rupe2024a}. Taken together, these arguments suggest that emergence is inseparable from the problem of representation. The issue is not merely whether emergent behavior exists, but how one should represent the organizational regularities through which it becomes visible.

On this reading, topology matters because it supplies a mathematically explicit vocabulary for describing the \emph{shape} of emergent organization. Topological descriptors do not, by themselves, identify causal mechanism, and a loop or cavity is not yet an explanation. Nevertheless, such descriptors can register aspects of organization that are difficult to articulate in purely statistical or edge-based terms. Connected components capture segmentation and accessibility; one-dimensional cycles can register recurrent coordination or exclusion relations; higher-dimensional cavities can indicate collective constraints or missing higher-order closure; persistence distinguishes fleeting artifacts from stable mesoscale structure. In this sense, topology does not model emergence by replacing dynamics with geometry. Rather, it characterizes the shape of the spaces, states, and relational configurations within which emergent behavior unfolds.

The appeal of this viewpoint is visible in the growing range of applications. Horak, Maleti\'{c}, and Rajkovi\'{c} helped establish an early bridge between complex networks and persistent homology \cite{Horak2009a}. Petri and colleagues later used homological scaffolds to identify mesoscale structure in functional brain networks, while Sizemore and colleagues showed that cliques and cavities in the human connectome reveal organization that is not reducible to ordinary graph summaries \cite{Petri2014a,Sizemore2018a}. Gidea and Katz used persistence landscapes to detect reorganization in financial time series around crash periods, and Syed Musa and colleagues showed that persistent homology can augment early-warning analysis for critical transitions in environmental systems \cite{Gidea2018a,Musa2021a}. Li and colleagues, in a rather different register, demonstrated that persistent homology can demarcate biologically meaningful morphospaces in plant morphology \cite{Li2018a}. What unifies these otherwise disparate applications is not a single domain or data type, but a shared intuition that organization across scale is itself an empirical signal.

At the same time, the contemporary literature remains fragmented. Work on persistent homology, higher-order networks, topological neuroscience, and topological indicators of transition has often advanced in parallel, even when these disparate bodies of literature are responding to a common underlying problem \cite{Battiston2020a,Bick2023a,Petri2014a,Gidea2018a}. The present review is organized around that common problem. Rather than treating topological data analysis as a disconnected toolbox, we argue that topological methods gain coherence when they are understood as a language for emergent organization for complex systems. This claim is deliberately stronger than the observation that topology has many applications, but weaker than the claim that topology should displace statistical, graph-theoretic, or dynamical approaches. Topology is most useful here as a representational framework that complements those traditions by preserving forms of multiscale and higher-order organization that they often compress away \cite{Torres2021a,Wasserman2018a}.

\paragraph{Scope and review strategy.}
This article is a selective narrative review rather than a systematic review, meta-analysis, or bibliometric survey. Its aim is not exhaustive coverage of every topological technique or every complex-systems application, but a conceptually organized synthesis of the literature most relevant to a specific claim: that topology is especially useful when the scientific problem concerns emergent organization, understood operationally as distributed, multiscale, and higher-order structure that is only partially visible in scalar or dyadic summaries. The literature discussed below was therefore chosen to serve three purposes: first, to identify canonical theoretical foundations for persistent and related topological methods; second, to examine methodological work on stability, inference, representation, and higher-order modeling; and third, to compare application domains in which topological observables have been used to detect reorganization, regime structure, transition, or early warning. This selection principle makes the review deliberately interpretive rather than exhaustive. Some technically important branches of topological data analysis are mentioned only briefly when they do not bear directly on these representational questions, and very recent frontier results are used illustratively rather than as settled consensus.

The sections that follow develop this argument in stages. The next section clarifies what counts as a ``topological method'' in the present review and distinguishes topological invariants from topology-inspired higher-order representations. Subsequent sections examine why multiscale persistence provides a principled way to separate robust structure from noise, why pairwise graphs are often insufficient for representing collective interactions, and why topology is especially valuable when the scientific problem centers on organization rather than magnitude. Later sections then turn to applications in regime-shift detection, anomaly identification, neuroscience, finance, biology, and engineered socio-technical systems. The central claim is that topology's practical success in structural diagnostics and early warning follows from a deeper representational virtue: it captures patterns of organization that are distributed, multiscale, and only partially visible in conventional descriptions of complex systems.

\section{What Counts as a ``Topological Method''?}
\label{sec:what_counts_topological_method}

The phrase ``topological method'' is elastic enough in contemporary interdisciplinary work that, unless it is defined carefully, it can collapse very different objects---persistence diagrams, Mapper graphs, simplicial complexes, and hypergraphs---into a single and misleading category. Carlsson, Wasserman, and Chazal and Michel each describe topological data analysis as a program in which one constructs topological spaces or filtered objects from data and then studies qualitative features that are stable under controlled deformation \cite{Carlsson2009a,Wasserman2018a,Chazal2021a}. For a review centered on complex systems, however, a purely narrow definition is not quite enough, because some of the most relevant literature is not only about invariant extraction; it is also about representational choice, especially when higher-order organization would dissolve in a pairwise graph \cite{Torres2021a,Battiston2020a,Bick2023a}.

\begin{quote}
In this review, a method counts as \emph{topological} in the narrow sense when it computes a topological invariant---or a stable transform of such an invariant---from a filtered topological object constructed from data. A method counts as \emph{topology-inspired} in the broader sense when it represents a system by simplicial, nerve, or other higher-order constructions that preserve qualitative relational organization even before an invariant is computed.
\end{quote}

This nested definition is important because it separates three families of work that are often conflated in practice: filtration-based invariant methods, Mapper/Reeb-type organizational summaries, and higher-order representational formalisms. In all three cases, the crucial point is that topology does not operate directly on an undifferentiated dataset. Some modeling step intervenes between observation and inference. At a sufficiently abstract level, the pipeline can be written as
\[
\text{data}
\;\longrightarrow\;
\text{topological object}
\;\longrightarrow\;
\text{filtration or cover}
\;\longrightarrow\;
\text{invariant or organizational summary}.
\]
This schematic pipeline is deliberately broad, but it captures the common logic of the methods relevant to the present review: one first decides how to turn the system into a space, complex, or cover, and only then asks which qualitative structures are robust enough to interpret.

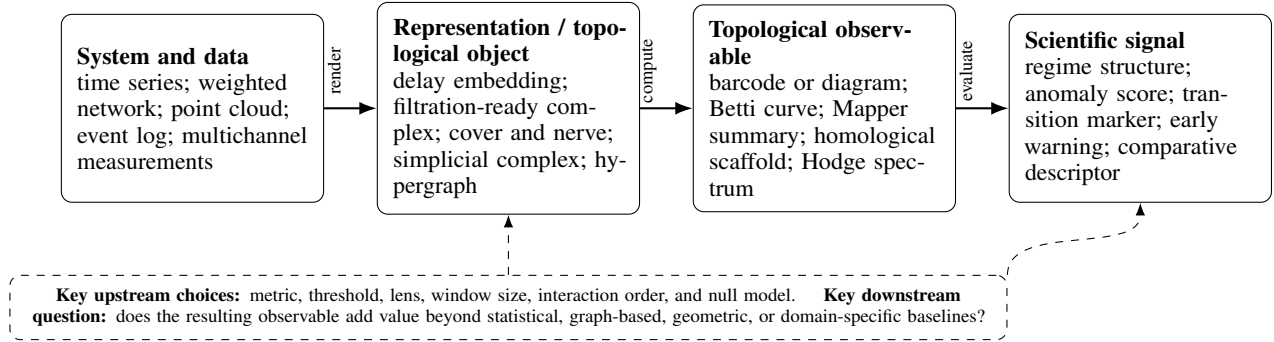
\begin{figure}[t]
\centering
\begin{tikzpicture}[
  >=Latex,
  stage/.style={
    draw,
    rounded corners,
    align=left,
    font=\small,
    text width=0.19\textwidth,
    minimum height=2.5cm,
    inner sep=6pt
  },
  note/.style={font=\scriptsize, align=center}
]

\node[stage] (data) {\textbf{System and data}\\
time series; weighted network; point cloud; event log; multichannel measurements};

\node[stage, right=7mm of data] (object) {\textbf{Representation / topological object}\\
delay embedding; filtration-ready complex; cover and nerve; simplicial complex; hypergraph};

\node[stage, right=7mm of object] (observable) {\textbf{Topological observable}\\
barcode or diagram; Betti curve; Mapper summary; homological scaffold; Hodge spectrum};

\node[stage, right=7mm of observable] (signal) {\textbf{Scientific signal}\\
regime structure; anomaly score; transition marker; early warning; comparative descriptor};

\draw[->, thick] (data) -- node[note, above, rotate=90, xshift=5mm]{render} (object);
\draw[->, thick] (object) -- node[note, above, rotate=90, xshift=5mm]{compute} (observable);
\draw[->, thick] (observable) -- node[note, above, rotate=90, xshift=5mm]{evaluate} (signal);

\node[
  draw,
  dashed,
  rounded corners,
  below=8mm of object,
  text width=0.80\textwidth,
  align=center,
  font=\scriptsize,
  inner sep=5pt
] (choices)
{\textbf{Key upstream choices:} metric, threshold, lens, window size, interaction order, and null model. \quad
 \textbf{Key downstream question:} does the resulting observable add value beyond statistical, graph-based, geometric, or domain-specific baselines?};

\draw[->, dashed] (choices.north) -- (object.south);
\draw[->, dashed] (choices.north east) .. controls +(0,0.7) and +(0,-0.7) .. (signal.south);

\end{tikzpicture}
\caption{Generic pipeline for topological analysis in complex systems. The main scientific judgment occurs upstream, where data are rendered as a topological or higher-order object; downstream observables are only as meaningful as those representational choices.}
\label{fig:topology-pipeline}
\end{figure}

Figure~\ref{fig:topology-pipeline} makes explicit the common analytic pipeline that recurs throughout the review, from raw system measurements to a topological signal that can be compared, monitored, or used for early warning.

\subsection{Persistent invariants and filtration-based constructions}

Persistent homology is the core instance of a topological method in the narrow sense. Zomorodian and Carlsson showed that the persistent homology of a filtered simplicial complex can be treated algebraically as the homology of a graded module, thereby supplying both the conceptual and algorithmic foundation for the subject \cite{Zomorodian2005a}. Edelsbrunner and Harer, Carlsson, and Chazal and Michel later presented this framework as a general recipe for extracting multiscale topological structure from data represented as point clouds, scalar fields, images, or weighted relational systems \cite{Edelsbrunner2008a,Carlsson2009a,Chazal2021a}. In its simplest form, one studies a nested sequence of spaces or complexes,
\[
K_0 \subseteq K_1 \subseteq \cdots \subseteq K_m,
\]
and asks how homological features evolve across the sequence. For each dimension \(k\), the associated homology groups \(H_k(K_i)\) describe \(k\)-dimensional holes, while the Betti numbers \(\beta_k\) record their ranks. Persistent homology tracks which homology classes are born, which survive, and which die as the filtration parameter changes \cite{Zomorodian2005a,Edelsbrunner2008a}.

The importance of the filtration step cannot be overstated. A point cloud may be converted into a Vietoris--Rips or \v{C}ech filtration; a scalar field may be analyzed through sublevel-set or cubical filtrations; a weighted graph may be turned into a clique or flag filtration; and a time series may be embedded into a reconstructed state space before any topological quantity is computed \cite{Chazal2021a,Adams2017a,Horak2009a}. In other words, persistent homology is never fully model-free. The topological information it extracts depends on how the analyst chooses to represent proximity, intensity, or interaction. This dependency should not be treated as a weakness peculiar to topology. Rather, it is the point at which domain knowledge enters the analysis in a mathematically explicit way.

The primary outputs of persistent homology are barcodes, persistence diagrams, Betti curves, and related summaries of birth and death across scale. These outputs are often discussed as if they were separate methods, but they are better understood as distinct representations of the same underlying invariant structure. Bubenik's persistence landscapes move persistence information into a Banach-space setting amenable to averaging and statistical inference, while Adams and colleagues' persistence images convert persistence diagrams into finite-dimensional vectors suitable for standard machine-learning pipelines \cite{Bubenik2015a,Adams2017a}. Mileyko, Mukherjee, and Harer supplied a metric-statistical framework for persistence diagrams using Wasserstein-type geometry, and Cohen-Steiner, Edelsbrunner, and Harer established a key stability theorem showing that small perturbations of the input induce controlled perturbations of the persistence diagram \cite{Mileyko2011a,CohenSteiner2007a}. The bottleneck distance highlights the largest unmatched discrepancy between diagrams, whereas \(p\)-Wasserstein distances aggregate many discrepancies at once; both are part of the now-standard metric apparatus of persistent homology \cite{Mileyko2011a,CohenSteiner2007a}. The crucial interpretive point is that persistence diagrams, landscapes, images, and distances are not rival notions of topology. They are alternate encodings or comparison tools for outputs that remain invariant-based.

For the purposes of this review, then, persistent homology and its derived summaries form the methodological core. They are the clearest examples of what it means to use topology as an inferential language rather than as a loose metaphor. They also illustrate a broader principle that will matter throughout the paper: topology becomes scientifically useful when qualitative structure is tracked across a controlled family of representations, rather than read off from a single threshold or snapshot.

\subsection{Mapper, nerves, and Reeb-type organizational summaries}

Mapper occupies a different methodological position. Singh, M\'emoli, and Carlsson introduced Mapper as a construction that produces a simplicial summary of data by pulling back a cover along a chosen filter (or lens) function and then clustering within the pullback sets \cite{Singh2007a}. In formal terms, Mapper is a nerve construction. In practical terms, it is often visualized through its 1-skeleton and discussed as a graph. Lum and colleagues helped popularize the method by showing that this cover-and-cluster pipeline can reveal coarse structure in complex, high-dimensional datasets that is difficult to recover from PCA or standard clustering alone \cite{Lum2013a}. Although users commonly speak of a ``Mapper graph,'' the underlying object is not merely graphical; it is combinatorial-topological, and its meaning depends on the nerve of overlapping sets rather than on adjacency alone \cite{Singh2007a,Lum2013a}.

Unlike persistent homology, however, Mapper is not naturally interpreted as an invariant of the data alone. Its output depends explicitly on the filter function, the cover resolution and overlap, the clustering rule, and the metric or similarity structure used in clustering. That dependence is not incidental. It is constitutive of what Mapper does. Carri\`ere and Oudot showed that one-dimensional Mapper can be analyzed as a discretized or ``pixelized'' approximation to a Reeb graph, and they established structure and stability results that make this relationship mathematically precise \cite{Carriere2018a}. Carri\`ere, Michel, and Oudot then extended this line of work by studying statistical convergence and parameter selection for Mapper, while Dey, M\'emoli, and Wang clarified the relation among nerves, Reeb spaces, Mapper, and multiscale Mapper \cite{Carriere2018b,Dey2017a}. Together, these papers make clear that Mapper should not be assimilated uncritically to persistent homology. It is not simply another persistence summary, nor is it merely an exploratory visualization device.

A more accurate characterization is that Mapper is a topological summary of organization relative to a chosen observable. The filter function determines which aspect of the data is made salient; the cover decides the scale at which local pieces are separated and reconnected; and the nerve records how those pieces overlap. For complex-systems analysis, this makes Mapper particularly attractive when the scientific question concerns branching, coarse segmentation, or regime structure in a state space or feature space. In such settings, the analyst may care less about counting \(k\)-dimensional holes in an invariant fashion than about recovering a structurally meaningful organization of states. Mapper therefore belongs inside the present review, but it belongs there for a different reason than persistent homology does. Persistent homology is primarily an invariant-producing method; Mapper is primarily an organizational summarizer grounded in topological constructions.

\subsection{Simplicial complexes, clique complexes, and hypergraphs}

Simplicial complexes, clique complexes, and hypergraphs occupy a boundary region between topological analysis and representational modeling. In one sense, they are merely the abstracta on which topological methods operate. In another sense, choosing to represent a system by such objects is already a substantive methodological commitment. Giusti, Ghrist, and Bassett emphasized this point in their review of higher-order neural data analysis: the dyadic assumption built into graph theory suppresses many forms of joint interaction, whereas simplicial complexes encode relations among three or more units directly \cite{Giusti2016a}. A simplicial complex is not just a graph with extra decoration. If a \(2\)-simplex is present, then all of its faces---the three edges and three vertices---must also be present. This ``downward closure'' property gives simplicial complexes a specifically topological character, because it imposes a structured hierarchy of relations across dimensions \cite{Giusti2016a,Bick2023a}.

Clique complexes and concurrence complexes are especially important in applied work. A clique complex is obtained by filling every complete subgraph of a graph with a simplex, thereby promoting all-to-all connectivity patterns into higher-dimensional objects. Horak, Maleti\'c, and Rajkovi\'c used this logic in an early and influential study of persistent homology in complex networks, showing how homological structure can be extracted from network data once the graph has been lifted to a simplicial complex \cite{Horak2009a}. Concurrence complexes proceed differently: instead of filling cliques in an existing graph, they encode directly observed co-activation or co-occurrence events as simplices \cite{Giusti2016a}. The difference matters because it separates two distinct modeling claims: one may either infer higher-order interaction from dense pairwise structure or measure it directly from joint observations.

Hypergraphs broaden the picture further by permitting higher-order interactions without downward closure. Battiston and colleagues, Bick and colleagues, and Zhang and colleagues have all stressed that hypergraphs and simplicial complexes are not interchangeable formalisms; they can encode different assumptions and generate different dynamics, even when both are said to represent ``higher-order interactions'' \cite{Battiston2020a,Bick2023a,Zhang2023a}. This distinction is conceptually important for the present review. Not every hypergraph method is automatically a topological method in the narrow sense. Hypergraphs become part of the topological story when they are used to preserve relational organization that ordinary graphs erase, when they serve as substrates for homological or filtration-based analysis, or when the contrast between hypergraphs and simplicial complexes clarifies what kind of higher-order structure is actually being claimed.

The resulting boundary is therefore deliberate. Simplicial complexes, clique complexes, and hypergraphs are included here not because every paper using them belongs to topological data analysis, but because complex-systems research increasingly uses them to make explicit claims about organization beyond pairwise structure. When persistence, homology, Hodge-type operators, or filtration-based reasoning are computed on these objects, they enter the narrow topological core. When the main contribution is representational---that is, when the move from edges to simplices or hyperedges is itself the scientific point---they remain topology-inspired but still central to the argument of this review.

This taxonomy prevents two recurring conflations. First, invariant-producing methods should not be confused with the objects on which those invariants are computed. Second, higher-order representations should not be collapsed into persistent homology merely because both draw on algebraic-topological language. The present review therefore has a core and a penumbra. The core consists of persistent homology, persistence diagrams and barcodes, and their stable statistical transforms. The penumbra consists of Mapper/Reeb-type summaries and higher-order representations whenever they preserve organizational structure that pairwise graphs and low-order statistics would erase. This scoped usage is narrow enough to remain rigorous, but broad enough to support the central claim of the paper: topology matters in complex systems because it preserves qualitative organization across scale and order.

\section{Multiscale Structure and Persistence}
\label{sec:multiscale_structure}

One of the defining difficulties of complex-systems analysis is that there is rarely a single privileged resolution at which the system becomes fully intelligible. Local interactions, mesoscale motifs, and global constraints can each be scientifically important, and their relative importance often changes with the question being asked. Torres and colleagues argue that this is fundamentally a problem of representation: a representation does not merely record a system but selects which structures are even available for analysis \cite{Torres2021a}. Clark and colleagues make a related point in the broader multiscale-modeling literature, describing the scientific challenge as one of traversing length and time scales without losing the effective structure needed for explanation and prediction \cite{Clark2021a}. Persistent topology is valuable in this setting because it replaces a single, fixed representation with a \emph{controlled family} of representations. Instead of asking whether a loop, cavity, or connected component exists at one arbitrarily chosen threshold, it asks which such features remain visible as the relevant notion of scale is varied.

\subsection{Filtrations as probes of scale}

The mathematical device that makes this possible is the filtration. Edelsbrunner, Letscher, and Zomorodian introduced topological persistence precisely to distinguish structural features from topological noise within a growing family of spaces \cite{Edelsbrunner2002a}. Carlsson, and Edelsbrunner and Harer, later presented this idea as one of the central conceptual advances of topological data analysis: rather than extracting topology from a single complex, one studies a nested sequence
\[
K_{\alpha_0} \subseteq K_{\alpha_1} \subseteq \cdots \subseteq K_{\alpha_m},
\qquad \alpha_0 \leq \alpha_1 \leq \cdots \leq \alpha_m,
\]
and tracks how the induced homology groups
\[
H_k(K_{\alpha_0}) \longrightarrow H_k(K_{\alpha_1}) \longrightarrow \cdots \longrightarrow H_k(K_{\alpha_m})
\]
change across that sequence \cite{Carlsson2009a,Edelsbrunner2008a}. In this formulation, topology is not extracted from a static object. It is extracted from the \emph{variation} of an object under a specified change in resolution.

This point is more important for complex systems than it may initially appear. The filtration parameter is not always a geometric radius in the ordinary sense. Depending on the problem, it may be a proximity threshold in a point cloud, a density or function level in a scalar field, an edge-weight threshold in a network, a clique-expansion parameter, or even a diffusion time used to induce an effective geometry on a graph \cite{Edelsbrunner2008a,Wasserman2018a,Horak2009a,Tran2019a}. In time-dependent settings, the filtration may be built from evolving functional networks or reconstructed state spaces, so that topological structure is interrogated not just across geometric scale but across observational or dynamical resolution \cite{Stolz2017a}. The practical consequence is that persistent homology does not presuppose a single notion of scale. It provides a framework for declaring a scale parameter explicitly and then asking what structural information is stable with respect to that declaration.

Seen in this light, filtrations are better understood as \emph{probes} of scale than as merely technical input objects. They encode the analyst's hypothesis about how microscopic detail should be relaxed. A Vietoris--Rips filtration probes what happens as isolated points are thickened by increasing pairwise proximity. A sublevel-set filtration probes how the topology of a landscape changes as one raises an energy, density, or measurement threshold. A weighted-network filtration probes how relational structure reorganizes as weak ties are added or removed. A diffusion-based construction, such as the one used by Tran and colleagues, probes how the geometry induced by flow or random walk changes with the timescale of diffusion \cite{Tran2019a}. In each case, the filtration is not a nuisance parameter that must be fixed before analysis can begin. It is the formal device by which the analysis becomes multiscale.

\subsection{Persistence as a robustness criterion}

Persistent homology then asks which topological classes survive across a nontrivial portion of this filtration. If a class is born at parameter value \(b\) and dies at \(d\), its persistence is \(d-b\). This quantity is often visualized through barcodes or persistence diagrams, but the important conceptual point is that persistence is not merely a record of duration. It is a criterion of \emph{robustness relative to scale}. Edelsbrunner, Letscher, and Zomorodian's original formulation already suggested this interpretation by classifying topological changes as features or noise according to their lifetime within the filtration \cite{Edelsbrunner2002a}. Later expositions by Carlsson and Wasserman made the same idea central to the statistical and data-analytic interpretation of the subject \cite{Carlsson2009a,Wasserman2018a}.

The theoretical appeal of this viewpoint is strengthened by stability results. Cohen-Steiner, Edelsbrunner, and Harer showed that persistence diagrams vary in a controlled way under perturbations of the underlying filtration function, providing a rigorous explanation for why persistent features can be interpreted as stable rather than accidental \cite{CohenSteiner2007a}. Fasy and colleagues extended this logic in a statistical direction by deriving confidence sets for persistence diagrams, thereby giving a principled way to separate topological signal from topological noise rather than relying only on visual inspection of long and short bars \cite{Fasy2014a}. Chazal and colleagues then pushed the robustness question further by showing that direct distance-function methods can be extremely sensitive to outliers, and that smoother alternatives such as the distance-to-a-measure and kernel distance support more robust topological inference \cite{Chazal2018a}. Taken together, these results justify a stronger claim than the informal slogan that ``long bars matter.'' They show that persistence can be treated as an explicit statistical and geometric notion of stability under controlled deformation.

At the same time, persistence should not be reified. A long interval in a barcode is not automatically a physically meaningful mechanism, nor is a short interval automatically irrelevant. Persistence is always \emph{relative} to a chosen filtration, and thus to a chosen notion of scale. Torres and colleagues' broader lesson about representation applies here as well: the method cannot determine on its own which scale variation is scientifically appropriate \cite{Torres2021a}. Persistent homology does not eliminate modeling judgment; it relocates that judgment to the choice of filtration. This is precisely why the method can be powerful in complex-systems work. Instead of hiding assumptions about scale in a single threshold choice, it forces those assumptions into the open.

Recent work by Turke{\v{s}}, Mont{\'u}far, and Otter is particularly useful on this point. They show that persistent homology performs especially well on tasks that are genuinely tied to structural or shape-related properties, and that it often remains competitive under limited data, restricted computational budgets, and noise \cite{Turkes2022a}. Their results are important not because they establish persistent homology as universally superior, but because they clarify the method's inductive bias. Persistent homology is strongest when the target phenomenon is expected to survive deformation, subsampling, or moderate corruption, and when the relevant information is organizational rather than purely metric. That conclusion fits naturally with the broader argument of this review: persistence is useful in complex systems because many complex-systems questions are about the stability of organization under changing resolution.

\subsection{Multiscale topology and the logic of coarse-graining}

The connection to complex-systems science becomes sharper when persistence is compared to coarse-graining. In multiscale modeling, one seeks reduced descriptions that suppress microscopic detail while preserving the variables or constraints that matter at larger scales. Clark and colleagues describe this as the central challenge of ``traversing scale,'' emphasizing that useful reduced models must remain faithful to the organizational structure of the underlying system even as they compress its degrees of freedom \cite{Clark2021a}. Persistent topology should not be identified with coarse-graining in a strict dynamical sense. It does not, by itself, derive effective equations of motion, eliminate fast variables, or implement a renormalization-group transformation. Nevertheless, there is a genuine family resemblance between these enterprises.

The resemblance lies in the question both approaches ask: \emph{what survives systematic changes in resolution?} In coarse-graining, the issue is which collective variables remain predictive after microscopic detail has been compressed. In persistent homology, the issue is which qualitative features remain visible as the scale parameter of a filtration changes. The analogy should not be overstated, but it is useful. Renormalization and other coarse-graining procedures are interested in invariants or effective quantities that remain stable under repeated transformation. Persistent homology is likewise interested in structural features that are not destroyed by moderate thickening, threshold adjustment, smoothing, or diffusion. Where coarse-graining compresses degrees of freedom, persistence compresses a family of topological changes into a summary of births, deaths, and stable intervals.

This is why persistent homology is better understood as a formalization of \emph{stability under deformation} than as a mere shape descriptor. The deformation in question is not always a literal geometric deformation. It may be a change in density threshold, an increase in allowed similarity, a diffusion process on a network, or a smoothing parameter applied to a noisy empirical measure. The point is that topology is not asked to identify every local irregularity. It is asked to identify the structures that remain legible when the analyst systematically relaxes microscopic detail. That is exactly the level at which many complex systems become scientifically interpretable: not at the scale of individual fluctuations, but at the scale where robust organization begins to emerge.

A useful way to phrase this is that persistent homology does not choose the correct scale for the analyst; rather, it organizes evidence \emph{across candidate scales}. This makes it especially attractive in settings where no single threshold is obviously privileged. Complex systems often present precisely this difficulty. A financial correlation network, a neural activity cloud, or a reconstructed attractor may all admit many plausible threshold choices, none of which can be defended \textit{a priori }as uniquely correct. Persistent topology converts that ambiguity into a feature of the analysis. Instead of selecting one resolution and hoping it is adequate, the method records how structure reorganizes across a range of resolutions and then asks which patterns remain stable.

\subsection{Consequences for complex-systems analysis}

Several strands of the literature illustrate this multiscale logic directly. Horak, Maleti\'{c}, and Rajkovi\'{c} showed early on that persistent homology can reveal nontrivial organization in complex networks once those networks are lifted into suitable filtrations or clique complexes \cite{Horak2009a}. Stolz, Harrington, and Porter extended this general approach to time-dependent functional networks, showing that evolving topological summaries can track changes in coupled dynamical systems that are not obvious from edge-level descriptions alone \cite{Stolz2017a}. Tran and colleagues then made scale itself the object of analysis by constructing diffusion-based point clouds at different diffusion times and using the resulting topological summaries as explicitly scale-variant descriptors of network structure \cite{Tran2019a}. In that framework, micro-, meso-, and macro-scale organization are not treated as separate problems; they are linked by a common topological pipeline.

Myers and colleagues provide a complementary example from time-series and dynamical-state detection. Rather than applying persistent homology directly to a time-delay embedding, they build coarse-grained state-space networks and then compute persistent homology on those reduced relational objects \cite{Myers2023a}. Their results are especially relevant for the present argument because they show two things at once. First, the topological signal can become clearer after an appropriate coarse-grained representation is constructed. Second, the success of the method depends on the fact that the coarse-graining preserves qualitative organization rather than only local numerical detail. In other words, the effectiveness of persistence there is not independent of representation. It depends on a representation that already respects the multiscale structure of the underlying dynamics.

This observation helps explain both the promise and the limits of persistent topology in complex-systems work. The promise is that persistent homology provides a principled way to study organization without forcing the analyst to commit prematurely to a single threshold or scale. The limit is that persistence does not absolve the analyst of choosing how scale should be represented. A poor filtration can obscure meaningful structure just as easily as a poor network construction or a poor statistical model can. Otter and colleagues underscore the practical side of this point by emphasizing that filtration choice and computational design strongly affect what can be extracted from real data \cite{Otter2017a}. Persistent homology is therefore not best understood as an automatic discovery device. It is a disciplined framework for asking whether a proposed representation preserves robust structural information across scale.

The main conclusion of this section is thus conceptual rather than merely technical. Persistent homology matters for complex systems not simply because it counts holes, but because it operationalizes the idea that structure should be tested for \emph{stability across resolution}. This is why persistent homology is more than another feature-extraction routine. It provides a mathematically explicit answer to a central question in complex-systems science: which aspects of organization survive when one stops looking at the system in only one way? That question leads directly to the next section. Once one accepts that organization may persist across scale, one must still decide what kinds of relations a representation ought to preserve. In many complex systems, pairwise graphs are not enough. The next step, therefore, is to ask how topology handles \emph{higher-order} organization beyond dyadic interaction.

\section{Higher-Order Structure Beyond Pairwise Interactions}
\label{sec:higher_order_structure}

Graph-based modeling has become so central to complex-systems science that pairwise interaction is often treated as the natural ontology of relational data. In many settings, that assumption is entirely appropriate. If the phenomenon of interest is adequately described by dyadic influence, pairwise transport, or binary adjacency, then graphs remain among the most powerful and interpretable tools available. The problem is that, in a growing range of systems, the graph is not merely incomplete but structurally misleading. Battiston and colleagues, Bick and colleagues, and Boccaletti and collaborators all emphasize that many biological, social, and technological systems exhibit interactions among three or more units that cannot be represented faithfully by dyadic edges without changing the effective system description \cite{Battiston2020a,Battiston2021a,Bick2023a,Boccaletti2023a}. When the interaction itself is polyadic, reducing it to a collection of pairwise relations can erase the order of the interaction, the overlap structure among groups, and the collective constraints that arise only at the level of the group.

This issue is not merely terminological. It goes to the heart of how complex systems are represented. Torres and colleagues argue that representational choices determine which observables, hypotheses, and explanatory strategies become available \cite{Torres2021a}. In the present context, this means that a graph, a hypergraph, and a simplicial complex are not interchangeable encodings of the same relational fact. They preserve different kinds of organization and therefore support different kinds of inference. The central claim of this section is that topology provides a natural language for higher-order structure because it does more than record that several units are related. It records how those relations are nested, how they close, and how they generate collective constraints across dimensions.

\subsection{Translating from higher-order patterns to higher-order relations}

A first clarification is essential. The phrase ``higher-order structure'' is used in the literature in more than one sense. Benson, Gleich, and Leskovec use it to refer to motif-level organization within an ordinary graph, showing that networks can possess rich organization at the level of small subgraphs even when the primitive relational object remains the edge \cite{Benson2016a}. This is an important and genuinely higher-order perspective, because motifs such as feed-forward loops, wedges, and triangles reveal organizational patterns not visible from degree counts alone. Yet motif-level higher-order organization inside a graph is not the same as a direct many-body interaction. A motif is a pattern \emph{of} dyadic links. A polyadic relation is a relation whose basic unit already involves a set of size greater than two.

That distinction matters because the two notions answer different scientific questions. Motif analysis asks how pairwise relations arrange themselves into recurrent patterns. Higher-order relational modeling asks whether the system is more faithfully represented by objects whose primitive units are groups rather than edges. Bick and colleagues explicitly note that the current literature mixes several research programs under the umbrella of higher-order networks, including topology and geometry of data, relational-data modeling, and dynamical systems with non-pairwise couplings \cite{Bick2023a}. For the purposes of this review, the decisive point is that many complex systems require the second and third of these senses. Meetings, coauthorship groups, biochemical reactions, synchronous neural assemblies, and collective reinforcement processes are often not well described by decomposing a group event into all of its constituent pairs. The issue is not that a pairwise projection cannot be formed. It can. The issue is that the projection can destroy the very structure one wants to explain.

A simple example illustrates the problem. Suppose three individuals jointly influence a fourth to adopt a new behavior. A pairwise projection can preserve the fact that the target individual is connected to each member of the triad, but it does not preserve whether the influence was exerted through three independent dyadic exposures or through a genuinely collective event. If the mechanism depends on reinforcement or co-presence, then the order of the interaction is causally relevant. Iacopini and colleagues make this point explicitly in their simplicial contagion model, where group interactions are treated as irreducible ingredients of the spreading process rather than as accumulated dyadic exposures \cite{Iacopini2019a}. In such cases, the graph is not wrong in a trivial sense; it is wrong because it erases the level at which the mechanism operates.

Recent review work further sharpens this point by noting that higher-order interactions need not always be directly observed. Battiston and colleagues argue that non-pairwise terms can also arise as \emph{effective} interactions in reduced descriptions of coupled dynamical systems, even when the microscopic coupling is pairwise \cite{Battiston2026a}. This is conceptually important. Higher-order structure is therefore not only a matter of data format; it can also be a matter of the correct coarse-grained model. In other words, the case for higher-order representation does not depend on naive realism about group interactions in raw data. It depends on whether the scientific level of description at which one seeks explanation is intrinsically supra-dyadic.

\subsection{Hypergraphs, simplicial complexes, and induced closure}

Once one accepts that dyadic graphs may be insufficient, the next question is which richer object should replace them. The two most prominent candidates are hypergraphs and simplicial complexes. A hypergraph \(\mathcal{H}=(V,E)\) consists of a vertex set \(V\) together with a collection of hyperedges \(e \subseteq V\), where each hyperedge may involve any number of vertices. A simplicial complex \(K\) is more structured: it is a family of simplices closed under inclusion, so that whenever \(\sigma \in K\) and \(\tau \subseteq \sigma\), one also has \(\tau \in K\) \cite{Giusti2016a,Bick2023a}. This downward-closure condition encodes a substantive assumption about relational organization.

The difference can be stated plainly. Hypergraphs represent group relations without requiring that all subgroups also count as relations of the same object. Simplicial complexes require exactly that closure. In a simplicial complex, a filled triangle is not just a set of three pairwise edges; it is a 2-simplex whose three edges and three vertices are automatically present as faces. Iacopini and colleagues emphasize that this feature makes simplicial complexes a special kind of hypergraph, appropriate when group interactions come with a nested hierarchy of lower-order relations \cite{Iacopini2019a}. Bick and colleagues stress the reciprocal point: hypergraphs and simplicial complexes are not interchangeable mathematical devices for saying ``more than two.'' They embody different assumptions about what counts as relational structure \cite{Bick2023a}.

This distinction becomes topologically meaningful in the simplest nontrivial case. In an ordinary graph, a triangle is also a cycle of length three. In a simplicial complex, those two situations can be separated. One may have three edges that form an unfilled 1-cycle, or one may add the 2-simplex that fills that cycle. The first case preserves a one-dimensional hole; the second destroys it. Topology therefore distinguishes mere circular adjacency from collective closure. This is one of the clearest reasons simplicial complexes matter for complex-systems analysis. They do not only encode that several pairwise relations coexist. They encode whether those relations close into a higher-order whole.

Salnikov, Cassese, and Lambiotte argue that this is precisely why simplicial complexes are useful in complex-systems research: they generalize network tools by allowing many-body interactions while simultaneously supplying a topological framework in which holes and cavities become meaningful observables \cite{Salnikov2018a}. Giusti, Ghrist, and Bassett make a closely related point for neural data, where concurrence among multiple units is often more naturally represented by simplices than by graphs \cite{Giusti2016a}. The important consequence is that higher-order representation is not simply about adding larger relational containers. It is about preserving the hierarchy of faces, closures, and absences that define collective organization.

At the same time, not all higher-order structure is directly observed. In many applications, one constructs a simplicial complex from lower-order data. Clique or flag complexes, for example, fill complete subgraphs with simplices, thereby lifting dense pairwise patterns into higher-dimensional objects. Horak, Maleti\'{c}, and Rajkovi\'{c} used this strategy in early work on persistent homology for complex networks \cite{Horak2009a}. Such constructions are often scientifically fruitful, but they represent \emph{inferred} higher-order organization rather than directly measured group interaction. This distinction should be kept visible. A clique complex built from a correlation graph and a simplicial complex built from observed co-participation events are mathematically comparable, but they rest on different empirical commitments. The first interprets closure from dyadic density; the second records closure from observed joint occurrence.

From the perspective of this review, the key lesson is that topology enters already at the level of representation. Hypergraphs preserve arity. Simplicial complexes preserve arity together with closure. Clique and concurrence constructions preserve inferred or observed patterns of collective participation. In each case, what is gained is not just expressiveness in the informal sense, but the ability to distinguish open from closed relational structure, nested from unnested groups, and empty from filled higher-dimensional patterns. These are precisely the kinds of distinctions that pairwise projections compress away.

\subsection{Higher-order operators and topological signal spaces}

The advantage of simplicial representation becomes even more pronounced once one moves from static structure to operators and dynamics. Graph theory supplies a powerful operator language through the graph Laplacian, but that operator acts on signals defined on vertices. Many complex systems, however, carry meaningful information on relations themselves: flows along edges, cyclic currents, trajectories, disagreements, and other observables that are not naturally reducible to node values. This is where higher-order topological operators become useful.

Let \(B_k\) denote the boundary matrix mapping oriented \(k\)-simplices to their oriented \((k-1)\)-dimensional faces. The \(k\)-dimensional Hodge Laplacian is then
\[
L_k = B_k^{\top} B_k + B_{k+1} B_{k+1}^{\top}.
\]
When \(k=0\), this reduces to the familiar graph Laplacian on vertices. When \(k=1\), however, it acts on edge signals, combining a lower-adjacency term \(B_1^{\top}B_1\), which couples edges through shared vertices, with an upper-adjacency term \(B_2B_2^{\top}\), which couples edges through shared triangles \cite{Parzanchevski2017a,Schaub2020a}. The operator therefore registers not only whether edges meet, but whether they participate in a common higher-order cell.

This extension is not merely formal. Parzanchevski and Rosenthal show that random walks on simplicial complexes reflect higher-dimensional homology and spectral structure, extending familiar graph-theoretic concepts such as recurrence, amenability, and spectral gap to the simplicial setting \cite{Parzanchevski2017a}. Schaub and colleagues construct a normalized Hodge 1-Laplacian and connect it to random walks on edges, spectral embeddings, and edge-space analogues of PageRank, thereby showing that higher-order topological operators can support concrete data-analytic workflows rather than only abstract theory \cite{Schaub2020a}. Barbarossa and Sardellitti push the same idea in signal-processing language, arguing that graph signal processing should be understood as a special case of a broader topological signal processing framework on simplicial complexes \cite{Barbarossa2020a}. In that setting, signals may live on vertices, edges, or higher simplices, and the interplay among dimensions becomes part of the analysis itself.

What these developments contribute to complex-systems science is a way of making collective constraints mathematically explicit. In edge space, for example, the Hodge decomposition separates signals into gradient, curl, and harmonic components. The gradient component corresponds to potential-like flow, the curl component to local circulation around filled higher-order cells, and the harmonic component to global circulation tied to nontrivial topology \cite{Schaub2020a,Barbarossa2020a}. This is precisely the sort of decomposition that a purely dyadic node-space analysis cannot supply. When the scientifically relevant object is a flow, trajectory ensemble, or relation-level pattern, higher-order operators transform topology from a descriptive background into an operational calculus.

For the present review, this matters because it shows that topology is not only about counting holes after a representation has been chosen. The representation itself can be endowed with operators that make higher-order structure dynamically and analytically tractable. The language of boundaries, co-boundaries, and Hodge Laplacians turns cycles, cavities, and closure relations into quantities that can interact with signal processing, diffusion, and stability analysis. Higher-order topology is therefore not an add-on to network science; it is an extension of the very space on which dynamics can be defined.

Table~\ref{tab:method-taxonomy} summarizes the distinct methodological roles of the main families discussed so far, with special attention to the difference between invariant-producing methods, organizational summaries, representational formalisms, and operator-based approaches.

\begin{table}[t]
\centering
\small
\setlength{\tabcolsep}{4pt}
\renewcommand{\arraystretch}{1.15}
\begin{tabularx}{\textwidth}{L{0.14\textwidth} L{0.18\textwidth} L{0.17\textwidth} Y Y}
\toprule
\textbf{Method family} &
\textbf{Primary mathematical object} &
\textbf{Methodological role} &
\textbf{Best suited for} &
\textbf{Main dependencies and risks} \\
\midrule

Persistent homology and derived summaries &
Filtered simplicial or cubical complex built from a point cloud, scalar field, image, or weighted network &
Invariant-based multiscale summary &
Robust connectedness, loops, cavities, and structural change across scale &
Depends on metric, filtration, windowing, and null model; stable answers can still answer the wrong representational question \\

Mapper and Reeb-type summaries &
Cover of a filter/lens space together with local clustering and a nerve construction &
Organizational summary rather than invariant &
Branching, regime structure, coarse state-space segmentation, exploratory state organization &
Sensitive to lens choice, cover resolution and overlap, clustering rule, and metric \\

Simplicial, clique, and concurrence complexes &
Higher-order relational object with downward closure &
Representation of nested group structure on which summaries or operators can be defined &
Collective interactions, closure, cavities, and higher-order organization beyond dyads &
Filled simplices may be observed or inferred; clique lifting can overstate higher-order structure \\

Hypergraphs &
Higher-order relational object without downward closure &
Representation of polyadic interaction when closure should not be assumed &
Many-body relations, arity-sensitive dynamics, and comparison with simplicial models &
Less directly topological unless transformed; conclusions depend on whether closure is scientifically justified \\

Hodge and operator-based methods on complexes &
Boundary operators, cochains, and higher-order Laplacians on a chosen complex &
Operator-level analysis of flows, cycles, harmonics, and dynamics on higher-order domains &
Signals on edges and simplices, diffusion, random walks, and dynamical decomposition &
Interpretation depends on orientation, complex construction, and the empirical meaning of higher-order cells \\

\bottomrule
\end{tabularx}
\caption{Taxonomy of the main method classes discussed in the review. The rows belong together because they preserve organizational structure beyond scalar or dyadic summaries, but they do not all contribute the same kind of evidence.}
\label{tab:method-taxonomy}
\end{table}

\subsection{Dynamical consequences of higher-order structure}

The representational differences just described would be interesting even if they had little effect on system behavior. The literature now shows the opposite. Higher-order structure changes collective dynamics in substantive ways. Battiston and colleagues' early reviews and their recent synthesis in \emph{Nature Reviews Physics} argue that non-pairwise interactions can generate dynamical phenomena that are absent from, or at least not generic in, pairwise models, including explosive synchronization, multistability, altered stability boundaries, and new dynamical states \cite{Battiston2020a,Battiston2026a}. The effect is not only quantitative, it is often qualitative.

Contagion models provide one of the clearest demonstrations. In their simplicial model of social contagion, Iacopini and colleagues show that group interactions can induce a discontinuous transition together with a bistable region in which healthy and endemic states coexist \cite{Iacopini2019a}. The mechanism is not equivalent to repeating pairwise influence more quickly. It depends on the existence of genuine group interactions. Ferraz de Arruda, Petri, and Moreno then showed that hypergraph-based contagion models likewise generate bistability and hysteresis, while also clarifying that the form of the higher-order representation affects the analytical treatment of the process \cite{deArruda2020a}. More recent reviews by Ferraz de Arruda, Aleta, and Moreno synthesize this literature and emphasize that higher-order structure reshapes thresholds, spreading pathways, and phase-transition behavior in contagion processes \cite{deArruda2024a}.

Synchronization provides a second major example. Gambuzza and colleagues develop a master-stability-style framework for simplicial complexes and show that the stability of synchronization depends on the simplicial organization of the system, not only on the graph backbone obtained by retaining pairwise edges \cite{Gambuzza2021a}. Their analysis makes especially clear that higher-order interactions do not merely perturb a graph model; they alter the operator structure that governs linear stability. Gallo and collaborators extend this further by studying directed higher-order interactions, showing that nonreciprocal group structure can induce or destroy synchronization in ways with no immediate dyadic analogue \cite{Gallo2022a}. Zhang, Lucas, and Battiston sharpen the representational lesson by showing that hypergraphs and simplicial complexes can shape collective dynamics differently, so that even among higher-order formalisms the choice of representation is scientifically consequential \cite{Zhang2023a}.

This last point deserves emphasis because it cuts against a common simplification in interdisciplinary work. Once one accepts that pairwise graphs are insufficient, it is tempting to treat all higher-order formalisms as roughly equivalent upgrades. The literature suggests otherwise. Hypergraphs, simplicial complexes, motif-based graph lifts, and effective higher-order couplings may all be useful, but they preserve different aspects of structure and therefore support different dynamics \cite{Bick2023a,Zhang2023a,Battiston2026a}. Representation is not an afterthought. It is the site at which structural assumptions become dynamical commitments.

From the perspective of the present review, the broader conclusion is straightforward. Topology provides a natural language for higher-order interactions not because every complex system should be rewritten as a simplicial complex, but because higher-order systems are often defined by closure relations, group constraints, and multilevel dependencies that pairwise models erase. Graphs tell us who is linked to whom. Higher-order topological representations can also tell us who acts together, which lower-order relations are implied by a collective interaction, whether a cycle is filled or open, and which relation-level signals are constrained by global structure. These distinctions are often the very conditions under which emergent organization becomes visible.

This observation prepares the ground for the next section. If topology is useful because it preserves kinds of organization that dyadic graphs and low-order summaries tend to flatten, then its relation to competing representational frameworks must be examined directly. The next task, therefore, is to ask what topology contributes that statistics, graph theory, and geometry do not already provide in the analysis of complex systems.

\section{Topology versus Competing Representations}
\label{sec:topology_competing_representations}

The case for topology in complex-systems science becomes clearest when it is compared, not to a caricature of existing methods, but to the representational traditions with which it actually coexists. Torres and colleagues argue that no representation is universally adequate for all scientific purposes; the choice of representation determines which structures are visible, which observables can be defined, and which explanations become tractable \cite{Torres2021a}. The question, then, is not whether topology should replace statistics, graph theory, or geometry in the abstract. It is which kinds of organization each of these traditions is best equipped to preserve. The argument of this section is that topology contributes something distinctive precisely because it is tuned to qualitative organization across scale and order. Statistics excels at uncertainty quantification and inference, graph theory at pairwise relational structure, and geometry at metric form and smooth variation. Topology becomes especially valuable when the scientific object is neither merely distributional, nor merely dyadic, nor merely geometric, but instead concerns robust organization that survives deformation, threshold variation, or changes in observational scale.

\subsection{Topology and statistics}

A comparison with statistics is the least straightforward, because topology is not external to statistics in any simple sense. Wasserman's review is especially useful on this point: he places topological data analysis squarely within the broader family of data-analytic methods for finding structure in data, alongside clustering, manifold estimation, mode estimation, and related nonparametric tools \cite{Wasserman2018a}. Chazal and Michel make a similar point when they describe TDA as a set of topological and geometric tools that can be used independently or in combination with data-analysis and statistical-learning methods \cite{Chazal2021a}. The relevant contrast, therefore, is not between ``topology'' and ``statistics'' as mutually exclusive enterprises. It is between different \emph{kinds} of summaries and inferential targets.

Classical statistical summaries often emphasize magnitudes, distributions, moments, regression relations, and dependence structures. These summaries are indispensable when the scientific question concerns average behavior, effect size, uncertainty, or model comparison under explicit assumptions. In such settings, topological summaries are not natural replacements. A persistence diagram does not directly substitute for an estimator of a mean effect, a confidence interval for a regression coefficient, or a causal estimand. Nor should it. What topology offers instead is a family of descriptors for \emph{organization}: connectedness, separation, loops, cavities, branching, and other global structures that are not generally reducible to a finite collection of low-order moments or pairwise associations \cite{Carlsson2009a,Wasserman2018a}. In that sense, topology enters where the scientific burden falls on relational form rather than on scalar magnitude alone.

This distinction matters in complex systems because many of the most interesting questions concern structural organization under uncertainty. A critical transition, a reconfiguration of brain activity, or the onset of coordinated market stress may not be signaled first by a dramatic shift in means or variances. It may instead appear as a change in the organization of a state space, correlation structure, or interaction pattern. Topological summaries are attractive in such settings because they are designed to capture qualitative configuration rather than only local quantitative fluctuation \cite{Carlsson2009a,Merelli2015a}. The point is not that topology is ``beyond'' statistics. It is that topology enlarges the class of structures that can be treated as statistical objects.

The modern TDA literature has, in fact, moved steadily in that direction. Bubenik's persistence landscapes embed persistence diagrams in a functional-analytic setting amenable to averaging and statistical comparison, while Adams and colleagues' persistence images provide finite-dimensional vector representations suitable for standard machine-learning pipelines \cite{Bubenik2015a,Adams2017a}. Fasy and colleagues construct confidence sets for persistence diagrams, thereby showing that topological signal can be addressed with inferential tools rather than by visual inspection alone \cite{Fasy2014a}. Chazal and collaborators push this program further by showing that robust topological inference often requires replacing naive distance-function constructions with more stable objects such as distance-to-a-measure or kernel distance \cite{Chazal2018a}. These developments are important conceptually because they show that the right contrast is not topology versus inference, but rather topology as a source of new inferential targets.

For present purposes, the deeper comparison is with the kinds of regularity each tradition privileges. Statistics is often strongest when variation can be summarized through distributions, likelihoods, or low-dimensional estimands. Topology is strongest when the relevant regularity is organizational and multiscale: when what matters is whether a system forms a coherent cycle, a branching structure, or a persistent cavity across a range of thresholds or resolutions \cite{Wasserman2018a,Chazal2021a}. This is why topology is often especially effective in settings where one does not wish to assume a parametric generative model but still wants a principled descriptor of global structure. In those cases, topology functions as a supplier of \emph{statistics of organization}.

At the same time, the limits of the comparison should remain clear. A topological descriptor is not automatically more informative than a classical statistical summary. If the scientific question concerns average treatment effects, variance decomposition, or optimal prediction under a well-specified probabilistic model, then topology may add little or nothing. The present review therefore treats topology not as a competitor to statistical methodology in general, but as a complementary source of observables for problems in which the object of interest is the configuration of relations rather than the value of a small set of parameters. That is precisely the sort of problem complex-systems science encounters repeatedly.

\subsection{Topology and graph theory}

The comparison with graph theory is more immediate because both graph-theoretic and topological approaches are often used to study relational data. Newman's classic survey established graph-based network science as one of the central frameworks for analyzing complex systems, emphasizing degree distributions, clustering, random graph models, growth, and dynamical processes on networks \cite{Newman2003a}. Bullmore and Sporns, and later Rubinov and Sporns, showed how graph-theoretic measures such as path length, efficiency, modularity, centrality, and resilience can turn large relational datasets into compact and interpretable summaries of network organization \cite{Bullmore2009a,Rubinov2010a}. When interactions are naturally dyadic and the scientific question concerns pairwise transport, shortest paths, hub structure, or community organization, graph theory is often exactly the right language.

Yet graph theory imposes a specific ontology: a system is represented by nodes and pairwise edges between them. In many complex systems that assumption is productive, but it is not neutral. It compresses joint interactions into dyadic relations and, in weighted settings, often requires thresholding or preprocessing choices before standard graph measures can be applied. Garrison and colleagues make this problem vivid in the context of functional brain networks, where graph measures and even the direction of group differences can vary substantially across thresholds \cite{Garrison2015a}. Their point generalizes well beyond neuroimaging. Whenever a rich weighted or continuous relational structure is reduced to a binary or sparsified graph, one must decide which edges count as present, which do not, and at what level of resolution the graph is supposed to live. Persistent topological methods are attractive partly because they turn that threshold ambiguity into an explicit part of the analysis rather than treating it as a preprocessing nuisance.

A second limitation is that graph measures are often interpreted as though they carried dynamical meaning independently of the dynamics under study. Curto and Morrison argue against that assumption in theoretical neuroscience, emphasizing that many graph-theoretic quantities are meaningful mainly for simple or linear dynamics and may become difficult to interpret once one moves to genuinely nonlinear systems \cite{Curto2019a}. Their critique is not a rejection of network science. It is a reminder that graph structure and system dynamics do not map onto one another in a universal way. The relevance of a degree distribution, a centrality measure, or a motif count depends on the mechanism one thinks the network is supporting. For complex systems more broadly, this is an important caution: graph theory offers a powerful vocabulary, but not a complete ontology of organization.

Topology contributes precisely where graph representations flatten structure that may be scientifically decisive. The simplest example is already familiar from the previous section. An unfilled triangle and a filled 2-simplex are identical as graphs, because both yield the same three vertices and three pairwise edges. Topologically, however, they are different objects. One contains a one-dimensional cycle; the other closes that cycle with a higher-order cell. Graph theory alone cannot record that distinction, because it has no native representation of filling, closure, or higher-dimensional cavity structure. This is why simplicial complexes, clique complexes, and other topology-inspired constructions matter: they preserve whether a pattern of pairwise adjacency constitutes mere local recurrence or a genuinely closed collective relation \cite{Giusti2016a,Bick2023a}.

Applications in neuroscience illustrate the gain. Petri and colleagues' homological scaffold of functional connectivity was designed precisely to recover mesoscale organization not exhausted by standard graph measures \cite{Petri2014a}. Sizemore and colleagues similarly showed that cliques and cavities in the human connectome reveal patterns of higher-order organization that are invisible if one restricts attention to edge-based summaries \cite{Sizemore2018a}. In these cases, topology does not simply add more descriptors to the graph-theoretic toolbox. It changes the ontological level at which structure is represented. Instead of asking only which nodes are linked, one can ask whether collections of relations close, whether cavities persist across threshold, and whether higher-order organizational motifs are robust across scale.

The distinction becomes even sharper in the higher-order network literature. Battiston and colleagues, Bick and colleagues, and Zhang and collaborators all show that hypergraphs and simplicial complexes are not merely elaborate graphs; they preserve group interactions and closure relations that can substantially alter collective dynamics \cite{Battiston2020a,Bick2023a,Zhang2023a}. In that sense, topology does not stand against graph theory so much as it extends the space of admissible relational objects beyond dyads. Graph theory remains indispensable, but topology is needed when the scientific object is not only the pattern of edges, but the organization of cycles, cavities, and higher-order closures that edges alone cannot encode.

It is therefore more accurate to say that topology asks a different question from graph theory. Graph-theoretic analysis often asks how a pairwise network is organized. Topological analysis asks what kinds of qualitative organization survive once one admits multiscale thresholding, higher-order closure, or deformation-invariant structure. In many complex-systems problems both questions are worth asking. The point is not to choose one vocabulary once and for all, but to recognize that graph-theoretic and topological descriptions preserve different families of structure.

\subsection{Topology and geometry}

The contrast with geometry is subtler still, because topology and geometry are historically and methodologically intertwined. Meilă and Zhang describe manifold learning as a family of methods that seeks low-dimensional structure in high-dimensional data and, in doing so, reveals the geometric shape of point clouds \cite{Meila2024a}. Geometry, in this broad data-analytic sense, is concerned with metric information: distances, neighborhood relations, curvature, smoothness, and embeddings. It is the natural language when the scientific problem depends on how far apart states are, which geodesics connect them, or whether data concentrate near a smooth low-dimensional manifold. Many complex systems generate exactly these kinds of questions, which is why manifold learning, spectral methods, and geometric inference have become so influential.

Topology shares some of this ambition but changes the level of description. Rather than preserving the full metric structure of an object, topology preserves features that survive continuous deformation: connectedness, separation, loops, cavities, and other coarse forms of organization. This does \emph{not} mean that topological methods are metric-free. Persistent homology depends crucially on a filtration, and filtrations are often built from metric or function-based information \cite{Carlsson2009a,Chazal2021a}. In practice, persistent homology is best understood not as an alternative to geometry, but as a disciplined procedure for extracting topological information \emph{from} geometric data. The filtration is the bridge. It turns distances, densities, weights, or directional function values into a family of spaces across which qualitative structure can be tracked.

The relationship between the two becomes especially clear in shape analysis. Turner, Mukherjee, and Boyer introduced the persistent homology transform as a representation of shapes and surfaces in which a collection of persistence diagrams serves as a shape statistic for computing distances and performing classification \cite{Turner2014a}. This is a paradigmatic example of topology working through geometry rather than against it. The input is geometric; the output is topological; and the result is a representation intended precisely for statistical comparison of shapes. Hiraoka and colleagues' analysis of amorphous solids makes a similar point in a different domain. Their persistence-diagram approach extracts hierarchical structures from atomic configurations, showing that topological summaries can reveal medium-range order that is difficult to describe through local geometric descriptors alone \cite{Hiraoka2016a}. In both cases, geometry provides the substrate, but topology supplies a language for the robust structural regularities that geometry by itself may not foreground.

This interplay is increasingly explicit in the study of neural representations as well. Lin and Kriegeskorte argue that understanding neural representations requires attention to both topology and geometry, not one at the expense of the other \cite{Lin2024a}. That observation is broadly instructive for complex systems. Geometry is often indispensable for understanding local deformation, similarity, and embedding. Topology becomes indispensable when the scientific question concerns the persistence of global organization under those deformations. The opposition between geometry and topology is therefore often misleading. In many of the most interesting applications, topology is what remains informative after the fine metric details have been deliberately compressed.

It is still useful, however, to state the difference sharply. Geometry asks questions such as: what are the relevant distances, local neighborhoods, and curvatures of the data? Topology asks: which structural relations remain after one stops caring about exact distances and begins caring about deformation-invariant organization? When the signal of interest is encoded in precise metric variation, geometry should lead. When the signal of interest is encoded in robust qualitative arrangement, topology should lead. Persistent topology is powerful precisely because it mediates between these regimes: it uses geometric or functional information to build filtrations, then records which topological structures survive as that information is varied \cite{Carlsson2009a,CohenSteiner2007a,Chazal2021a}.

\subsection{A division of representational labor}

The comparisons above suggest a division of labor rather than a competition for supremacy. Statistics is best understood as the language of uncertainty, estimation, and principled comparison under variability. Graph theory is the language of pairwise relational organization and network-level descriptors. Geometry is the language of metric form, embedding, and local shape. Topology is the language of robust qualitative organization across scale and, increasingly, across interaction order. None of these languages is sufficient for all complex-systems problems. Each reveals structures that the others compress.

This is also why topology is a plausible candidate for a language for emergent organization. Emergent organization is usually interesting precisely because it is neither exhausted by scalar summaries, nor reducible to dyadic adjacency, nor identical to fine metric detail. It concerns collective patterns that persist above the level of local fluctuation and that often become visible only when one looks at structure across scales. Topology is not unique in being able to say something about such patterns, but it is unusually well suited to preserving them once they have been represented appropriately. The next section develops that claim directly by asking not merely what topology is different from, but why it is so often drawn upon when the target of explanation is emergent organization itself.

\section{Topology as a Language for Emergent Organization}
\label{sec:topology_language_of_emergence}

The argument of the preceding sections can now be stated more precisely. Topology has become increasingly valuable in complex-systems science not simply because it offers a new family of descriptors, but because many complex-systems problems are, at bottom, problems of representing emergent organization. Simon's account of hierarchical organization, Anderson's claim that ``more is different,'' Ladyman and colleagues' attempt to characterize complex systems through robust organization, Crutchfield's treatment of emergence as a problem of model construction, and Rupe and Crutchfield's recent synthesis of emergent organization all point toward a shared difficulty: scientifically important structure often appears at a level that is not transparently visible from the raw states of individual components alone \cite{Simon1962a,Anderson1972a,Ladyman2013a,Crutchfield1994a,Rupe2024a}. The question is not merely whether large systems display surprising behavior. It is how one should represent that behavior once it becomes organized, stable, and explanatorily salient.

In the present review, \emph{emergence} is used in an operational rather than a metaphysically maximal sense. The term does not require an appeal to strong emergence, irreducible laws, or ontological novelty in the strict philosophical sense. What it does require is a conjunction of three features. First, the phenomenon must be \emph{distributed}: it arises from interactions among many units rather than being attributable to a single component. Second, it must exhibit \emph{macro-level organization}: there is some relatively stable pattern, constraint, or regularity at the system level. Third, and most important for the present argument, it must involve a \emph{representational gain}: the organization becomes visible only once the system is described in an appropriate macroscopic or relational language, rather than through a mere listing of microscopic values or low-order summaries \cite{Anderson1972a,Simon1962a,Crutchfield1994a,Torres2021a}. This third condition is crucial. It means that emergence is not only a property of the system; it is also a challenge for representation.

That representational challenge is precisely where topology becomes scientifically attractive. Topology does not offer a direct replacement for dynamical equations, statistical inference, or mechanistic explanation. What it offers is a vocabulary for describing organization at the level where emergent phenomena typically become legible: connectedness, branching, recurrence, closure, cavities, and persistence across scale \cite{Carlsson2009a,Wasserman2018a}. If one accepts, with Crutchfield, that the detection of emergence depends on the innovation of suitable model classes, then topology can be understood as one such model class \cite{Crutchfield1994a}. It does not resolve the philosophical debate over what emergence \emph{ultimately} is. It does, however, provide a disciplined way of representing structures that are distributed, multiscale, and qualitatively stable. That is enough to make it highly relevant to complex-systems science.

\subsection{Emergence in an operational sense}

The operational reading of emergence adopted here is deliberately modest, but it is not weak. It captures a family of scientific situations in which system-level structure matters more than isolated local values. A coherent attractor in a nonlinear dynamical system, a cell assembly in neural activity, a market-wide stress regime, or a phase transition in a many-body system are all emergent in this operational sense. In each case, the phenomenon is generated collectively, but it becomes scientifically tractable only after one introduces a representation in which the collective organization itself can be identified and compared \cite{Simon1962a,Anderson1972a,Ladyman2013a,Rupe2024a}.

This reading also clarifies why emergence should not be confused with mere complexity or mere unpredictability. A system may be complicated without exhibiting stable macroscopic organization, and it may be unpredictable without giving rise to a coherent large-scale pattern. The scientific interest of emergence lies not in surprise alone, but in the appearance of organized regularity above the level of individual components. Simon's emphasis on hierarchy and near-decomposability already suggested that complex systems become intelligible when their nested organization is made explicit \cite{Simon1962a}. Anderson's argument likewise implied that the appropriate concepts at one scale need not be reducible, in practice or in explanatory usefulness, to the concepts of a lower scale \cite{Anderson1972a}. Ladyman and colleagues later sharpened this point by treating robust organization and memory as central to many complex systems \cite{Ladyman2013a}. Rupe and Crutchfield extend this tradition by arguing that the identification of organization requires mathematically explicit structure, not just an intuitive sense that ``something new'' has appeared \cite{Rupe2024a}.

The present review adopts that last point as a guiding principle. Emergence becomes scientifically useful when it can be tied to observables that persist long enough, and clearly enough, to support analysis. Topology is relevant because it specializes in precisely this problem: how to describe qualitative structure that survives moderate deformation, threshold change, or coarse-graining. This does not mean that every emergent phenomenon is topological. It means that many emergent phenomena are best described at a level where topological structure becomes meaningful.

A further advantage of the operational reading is that it avoids an unnecessary opposition between emergence and mechanism. Topological descriptions do not replace mechanism; they organize the space within which mechanisms can be identified. A persistent loop in a reconstructed state space does not by itself explain an oscillation, nor does a cavity in a neural simplicial complex by itself explain cognition. What such descriptors do is identify the macroscopic organization that a mechanistic explanation must eventually account for. In that sense, topology is less a theory of emergence than a \emph{language} in which emergent organization can be written.

\subsection{Topological signatures of emergent organization}

The claim that topology can function as a language for emergent organization should not be interpreted too literally. Topological features do not carry universal semantic meanings independent of context. A one-dimensional cycle may indicate periodicity in a delay embedding, feedback or recurrent accessibility in a state-transition graph, or unfilled closure in a simplicial complex. A cavity may indicate a collective constraint in one domain and an exclusion region in another. Torres and colleagues' broader lesson about representational choice applies here as well: the scientific meaning of a topological feature depends on the object from which it is computed \cite{Torres2021a}. Topology is therefore not a dictionary in which each homology class translates automatically into one fixed kind of emergent behavior.

Even with that caution, several recurring correspondences appear across the literature. Connected components and branching structures are often the simplest signatures of emergent differentiation. When a system organizes into multiple coherent regimes, disconnected or weakly connected regions of a state space can register that separation at the level of representation. One-dimensional cycles often signal recurrent or oscillatory organization. In delay-coordinate or sliding-window embeddings, periodic and quasiperiodic signals naturally generate loop-like or toroidal structures, and persistent homology can quantify how strongly that recurrent organization is expressed \cite{Packard1980a,Takens1981a,Perea2015a}. Higher-dimensional cavities, by contrast, are often associated with collective closure or collective constraint. In simplicial or clique-complex settings, such cavities can encode patterns of coordination or exclusion that are not visible at the level of individual edges or pairwise correlations \cite{Petri2014a,Sizemore2018a,Giusti2015a,Curto2025a}.

What unifies these cases is not a single interpretation of holes. It is the fact that topological features describe \emph{organization of relations}. This is precisely the level at which many emergent phenomena appear. A periodic orbit is not a property of any one sampled point in a time series. A neural representation is not a property of any one neuron considered in isolation. A crisis regime in a market is not the property of any one asset return. In each case, the scientifically relevant object is a collective pattern of relations among states, variables, or units. Topology supplies descriptors for those patterns once the system has been represented in a space where such relations become explicit.

Persistence strengthens this connection to emergence. Emergent organization is rarely interesting because it appears at one arbitrarily chosen threshold. It is interesting because it remains identifiable across a nontrivial range of observational scales, thresholds, or perturbations. Persistent homology captures exactly this form of robustness. Carlsson and Edelsbrunner and Harer emphasized the multiscale nature of this framework, while Cohen-Steiner, Edelsbrunner, and Harer established stability results that justify treating persistent features as robust under controlled perturbation \cite{Carlsson2009a,Edelsbrunner2008a,CohenSteiner2007a}. This makes persistence particularly well suited to the study of emergence. A macroscopic pattern is scientifically meaningful not because it is visually striking at one scale, but because it survives the reasonable deformation of the representation. In that sense, persistence is a mathematical analogue of the robustness we normally expect from emergent organization.

It is worth stating the limitation directly. Topological signatures are not explanations of emergence in themselves. They do not establish causality, derive effective laws, or determine whether a given phenomenon is ``really'' emergent in a metaphysical sense. Their role is representational. They tell us where stable organization resides and how it changes across scale, threshold, or order. That role is already substantial. In many complex systems, the first scientific challenge is not to explain an emergent pattern mechanistically, but to identify it in a way that is robust, comparable, and faithful to the distributed nature of the underlying interactions. Topology is especially powerful at exactly that stage.

\subsection{Emergence across representations}

One reason the ``language for emergent organization'' formulation is useful is that it draws attention to a common logic across otherwise very different applications. The same topological vocabulary can be brought to bear on reconstructed dynamical state spaces, neural activity or representational manifolds, and market-level coordination structures, even though the underlying systems are physically and conceptually distinct. What carries across these cases is not the domain, but the representational situation: in each case, the emergent organization lives in a space of relations rather than in a list of isolated measurements.

\subsubsection{Time-delay embeddings and attractor topology}

The most classical example comes from nonlinear dynamics. Packard and colleagues showed that the geometry of a dynamical system can, under appropriate conditions, be reconstructed from a scalar time series using delay coordinates, and Takens gave the canonical embedding theorem that formalized this strategy \cite{Packard1980a,Takens1981a}. This result is foundational for the present argument because it reveals an important shift in where system-level structure resides. The raw observable is one-dimensional, but the emergent organization of the dynamics lives in a reconstructed state space. Periodicity, quasiperiodicity, and more complicated attractor structure are not obvious from a mere list of successive scalar measurements. They become visible only after one adopts a representation in which the collective organization of trajectories can appear.

Topological methods are especially natural in this setting because attractors are often better characterized by qualitative shape than by any single local statistic. Perea and Harer showed that sliding-window embeddings combined with persistent homology provide a principled way to quantify periodicity through the persistence of loop structure \cite{Perea2015a}. Mittal and Gupta later demonstrated that persistent homology can detect bifurcations and the onset of chaotic behavior in complex dynamical systems \cite{Mittal2017a}. In both cases, the relevant topological features are not decorative summaries layered on top of a finished dynamical analysis. They are the observables through which the emergent organization of the dynamics becomes legible. The topology of the reconstructed attractor is not the mechanism of the dynamics, but it is a natural macroscopic signature of the regime the system inhabits.

This is an important model for the broader review because it shows what a topological description of emergence looks like in its cleanest form. The micro-level signal is a sequence of measurements. The emergent object is a global organization in state space. The topological descriptor captures that organization precisely because it abstracts away from the exact coordinates while preserving connectedness, recurrence, and persistence. That pattern will recur in more complicated applied settings.

\subsubsection{Neural activity, functional organization, and representation}

Neuroscience provides a second, and in some ways deeper, illustration. Here the relevant emergent object is often not a dynamical attractor reconstructed from one variable, but a collective code, scaffold, or manifold generated by many interacting neurons or brain regions. Giusti and colleagues introduced clique topology as a way of detecting intrinsic structure in neural correlation matrices that is invariant under nonlinear monotone transformations, thereby addressing a problem that frequently undermines more standard matrix-analytic approaches \cite{Giusti2015a}. Petri and colleagues then used homological scaffolds to characterize mesoscale organization in functional brain networks, and Sizemore and colleagues showed that cliques and cavities in the human connectome reveal organization beyond ordinary edge-based network summaries \cite{Petri2014a,Sizemore2018a}. More recently, Lin and Kriegeskorte have argued that the topology and geometry of neural representations should be studied together, while Curto and Sanderson review a growing topological neuroscience literature aimed at linking circuits to function \cite{Lin2024a,Curto2025a}.

What makes these examples particularly relevant to emergence is that the scientifically interesting pattern is not localized in any single neuron, voxel, or pairwise edge. It lies in the organization of the ensemble. A neural representation is emergent in the operational sense used here because it is jointly produced, functionally significant, and visible only at a collective level of description. Topology contributes because it captures features of that collective organization that are robust to certain nonlinear distortions and threshold choices. In this context, holes and cavities should not be interpreted mystically. They are structural summaries of how neural activity or connectivity organizes itself in a high-dimensional relational space.

This has a further implication for the broader thesis of the paper. In the neural setting, topology does not merely reveal organization \emph{after} function has already been identified independently. It often helps formulate what counts as the relevant system-level structure in the first place. That is exactly what one would expect from a language for emergent organization. The role of the language is not only to describe known objects, but to make certain kinds of macroscopic object available for analysis.

\subsubsection{Markets and collective coordination structure}

Financial systems offer a third example in which the emergent object is unmistakably collective. The price of any individual asset is noisy, path-dependent, and heavily contingent. Yet market crises, regime shifts, and stress episodes are typically experienced and studied as system-level reorganizations of dependence and coordination. Gidea and Katz showed that persistent-landscape summaries of financial time series can reveal structural changes associated with major crash periods, while Ismail and colleagues used persistence-derived time series in conjunction with critical-slowing-down indicators to study early warning of financial crises \cite{Gidea2018a,Ismail2022b}. In these studies, topology is not used because markets are literally geometric objects in any everyday sense. It is used because market-wide organization emerges in the \emph{relations} among assets, windows, or derived configurations, and those relations can reorganize collectively before a crisis becomes obvious in low-order summaries alone.

This is a useful counterpoint to the dynamical and neural examples because it shows that the same topological logic applies even when the system is socio-technical and the underlying mechanism is heterogeneous. What emerges is not a single attractor or a single code, but a collective coordination structure: a changing pattern of alignment, concentration, and stress distributed across many assets. The topological descriptors are effective because they summarize the shape of that coordination structure rather than the trajectory of any one instrument. In this sense, topology functions as a language for emergent organization across domains precisely because it is not tied to the ontology of any one field. It operates at the level of organized relations.

\subsection{What topology can and cannot do for emergence}

The preceding discussion supports a strong but limited conclusion. Topology provides a language for emergent organization because emergent phenomena are often most visible as robust organization in spaces of relations, and topology is designed to describe such organization. It does this by encoding connectedness, recurrence, closure, cavity structure, and persistence across scale. These descriptors are especially valuable when the system-level regularity is distributed, multiscale, or only partially visible through pairwise or local summaries.

At the same time, topology does not settle the explanatory problem of emergence. It cannot, by itself, tell us which micro-mechanisms produce a cavity in a connectome, why a particular market configuration becomes unstable, or what causal process underlies a reconstructed attractor. It also does not guarantee that every interesting emergent phenomenon will have a clean topological signature. Some phenomena may be better captured statistically, geometrically, or through explicit mechanistic modeling. The claim of this review is therefore not that topology is the universal language for emergent organization, but that it is one of the most natural mathematical languages for emergent \emph{organization} in complex systems.

That distinction matters because it points directly toward the applied payoff. Once topological descriptors are understood as signatures of emergent organization, their use in anomaly detection, structural diagnostics, and early warning becomes much easier to interpret. They are valuable in those contexts not because they are fashionable descriptors of ``shape,'' but because regime shifts and anomalies in complex systems are often reorganizations of collective structure before they are simple excursions in scalar observables. The next section develops this practical consequence directly by examining topological methods for regime shifts, anomalies, and early warning signals in complex systems.

\section{From Representation to Signal: Applications in Detection and Early Warning}
\label{sec:detection_early_warning}

Before turning to applications, one distinction from Section~2 should remain explicit. The literature reviewed below does not provide one uniform kind of topological evidence. Persistent homology and its derived summaries contribute invariant-based observables computed from filtered objects; Mapper and related nerve constructions contribute organizational summaries relative to chosen filters, covers, and clustering rules; simplicial, hypergraph, and Hodge-type approaches often contribute representational or operator-level structure before any invariant is extracted. These families belong together in the present review not because they do the same methodological work, but because they address a shared representational problem: how to preserve distributed organization that would be weakened or erased by scalar summaries or dyadic graphs alone. The applications section should therefore be read comparatively. In some domains topology helps by stabilizing invariants across scale; in others it helps by choosing a higher-order object on which the relevant organization can be expressed at all.

The conceptual claim of the previous section has a direct applied consequence. If topology is useful because it captures emergent organization, then changes in topological organization should themselves become operational signals of instability, anomaly, or transition. Scheffer and colleagues helped establish the modern early-warning literature by emphasizing generic precursors such as rising variance and autocorrelation near critical transitions, while George and colleagues later reviewed the expansion of this literature toward multivariate, recurrence-based, network-based, and machine-learning indicators \cite{Scheffer2009a,George2023a}. The topological literature enters this conversation by asking a different but complementary question: can the changing \emph{shape} of a reconstructed state space, weighted network, or multivariate dependence structure serve as a warning signal in its own right?

The answer, across several domains, appears to be yes, but with an important qualification. Topological methods rarely produce warnings directly from raw observations. Instead, they transform structural reorganization into an observable that can be monitored over time: a Betti curve, a persistence-landscape norm, a Wasserstein trajectory, a Mapper-based transition map, or a topological distance between successive windows. In other words, topology typically functions as a \emph{representation-to-signal} device. The data are first turned into a state space, a correlation complex, a dynamic network, or another topological object; a persistent or higher-order summary is then extracted; and that summary is finally converted into a time-indexed statistic or change score. At a schematic level, the workflow is
\[
\text{observations}
\longrightarrow
\text{topological representation}
\longrightarrow
\text{topological observable}
\longrightarrow
\text{warning statistic or change score}.
\]

This applied literature is also heterogeneous in a way that should be acknowledged explicitly. Not every paper discussed below delivers \textit{ex ante} forecasting in the strict sense. Some studies genuinely aim at early warning before a transition occurs, while others are better described as structural detection, regime recognition, or state-space cartography. Treating them together is justified because they solve the same methodological problem: they convert changing organization into an operational signal. That common logic is clearer than the differences in domain.

\subsection{Detecting regime shifts in dynamical systems}

Nonlinear dynamical systems provide the cleanest test bed for topological warning methods because the transitions of interest are mathematically well defined. Bifurcations, crises, intermittency, and the onset of chaos all mark qualitative reorganization of the underlying dynamics. Classical tools such as Lyapunov exponents, Poincar\'e sections, spectral methods, and critical-slowing-down indicators remain indispensable, but they are not always easy to estimate from noisy, finite, or weakly observed data \cite{Scheffer2009a,George2023a}. Topological methods become attractive precisely because they target the changing organization of the reconstructed attractor rather than only scalar summary measures.

Following the delay-coordinate program initiated by Packard and colleagues and formalized by Takens, several studies first reconstruct a state space from a scalar time series and then apply persistent homology to sliding windows of the resulting point cloud \cite{Packard1980a,Takens1981a,Perea2015a}. In this setting, the topological signal is not attached to individual time points but to the evolving organization of neighborhoods, loops, and other persistent features across windows. Perea and Harer showed how this strategy can quantify periodicity, thereby making recurrence itself a topological observable rather than only a spectral or geometric one \cite{Perea2015a}. The general lesson is that regime change can be formulated as a change in the topology of the reconstructed state-space geometry.

Mittal and Gupta made this idea explicit by using persistent homology to characterize and detect bifurcations and chaos in canonical nonlinear systems \cite{Mittal2017a}. Their contribution is important because it frames topological summaries as regime-sensitive observables rather than as generic features for downstream classification. G\"uzel, Munch, and Khasawneh later proposed CROCKER plots as a topological summary for bifurcation analysis, showing that Betti-number trajectories across scale and control parameter can reveal transitions between periodic and chaotic behavior while retaining useful geometric information about the changing attractor \cite{Guzel2022a}. Dee Algar and colleagues pursued a related goal from a different angle by combining persistent homology with a transformation-cost metric between persistence diagrams to track dynamical regime change in both model systems and physiological time series \cite{DeeAlgar2021a}. Tanweer and colleagues then extended the basic logic to stochastic dynamical systems, where they used homological bifurcation plots to identify phenomenological bifurcations in probability-density structure rather than only deterministic attractor change \cite{Tanweer2024a}. More recently, Shah and colleagues used Betti-vector norms and persistence-landscape summaries to uncover transitions to chaos and to compare topological indicators with Lyapunov-based characterizations \cite{Shah2025a}.

A complementary line of work shifts attention from continuous attractors to relational reconstructions of state-space behavior. Myers, Munch, and Khasawneh showed that persistent homology of coarse-grained state-space networks can detect dynamic-state changes, illustrating that the topological signal can sometimes become clearer \emph{after} an appropriate relational representation is constructed \cite{Myers2019a}. This point matters because it reinforces the broader thesis of the review: topology is effective not because it is magically model free, but because it can preserve qualitative organization once a useful representation has been chosen.

Across these studies, the system is a time-dependent nonlinear process, the construction is typically a delay embedding or a derived network, the topological observable is a persistence summary or homology trajectory, and the signal is a time-varying score used to identify or anticipate regime change. Compared with scalar early-warning measures, topology has a specific advantage: it can respond to qualitative reorganization even when means, variances, or autocorrelations change only weakly. The tradeoff is that the resulting signal inherits sensitivity to embedding dimension, window length, sampling density, and filtration design. In dynamical systems, topology is therefore best viewed as a structural diagnostic of impending or ongoing regime change, not as a universal replacement for Lyapunov-based or critical-slowing-down methods.

\subsection{Financial systems and market stress}

Financial markets provide a more difficult and arguably more realistic environment for topological warning methods. The objects of interest are noisy, nonstationary, and high dimensional, and the transition of concern is usually not a clean bifurcation in one trajectory but a collective reorganization of dependence structure across many assets. This is precisely why the financial literature is so informative for the present review. It shows topology operating in a setting where emergent organization is relational, transient, and difficult to summarize with conventional univariate indicators alone.

One family of studies works directly with sliding windows of price or return time series. Gidea and Katz used persistence landscapes to study the topological structure of financial time series around crash periods, arguing that norms of the resulting landscapes can function as indicators of structural instability \cite{Gidea2018a}. Gidea and colleagues later extended this strategy to cryptocurrency markets, combining persistent homology with clustering to recognize critical transitions in highly erratic time series \cite{Gidea2020a}. Katz and Biem then developed a time-resolved topological framework for market instabilities in which sliding-window embeddings and persistence landscapes were used to identify transient loop structures associated with episodes of sector-wide stress \cite{Katz2021a}. In these studies, the topological signal emerges from the evolving geometry of short-horizon market dynamics rather than from a static network snapshot.

A second family begins with multivariate correlation structure rather than delay embeddings. Yen and Cheong applied TDA to stock markets in Singapore and Taiwan to study topological changes associated with crash periods \cite{Yen2021a}. Ismail and colleagues constructed persistence-based \(L^{1}\)-norm time series and then analyzed those signals alongside critical-slowing-down indicators to detect early-warning signals of historical financial crises in multiple markets \cite{Ismail2022a}. Their follow-up analysis across different correlation constructions is especially useful because it makes explicit how much of the signal depends on the underlying market representation rather than on persistent homology alone \cite{Ismail2022b}. That methodological transparency is valuable: it shows that topological early warning in finance is always a joint product of market construction and topological summary.

The structure of the application is therefore quite consistent. The system is a collection of interacting assets; the construction is either a sliding-window point cloud or a weighted correlation network; the topological observable is usually a persistence landscape, a diagram-based norm, or a related scalar summary; and the signal is a time series whose excursions, trends, or clusters are interpreted as evidence of mounting systemic stress. Compared with volatility and correlation spikes, topology targets the geometry of collective reorganization rather than only the magnitude of individual price moves. That is a genuine strength. A market may become more fragile because its dependence structure becomes more tightly coordinated or more abruptly reconfigurable even before univariate returns appear extraordinary.

At the same time, the evidentiary base in finance should not be overstated. Much of the literature remains retrospective and case-driven rather than benchmarked across a large number of crises. Warning horizons vary, false positives remain a concern, and the interpretability of specific topological features is often weaker than that of standard econometric variables. The strongest use of topology in finance is therefore not as a stand-alone crash oracle, but as a complementary structural indicator of systemic coordination, market stress, and dependence reorganization.

\subsection{Neuroscience and brain-state transitions}

Neuroscience presents a different version of the warning problem. The main challenge is often not forecasting a singular collapse, but detecting transitions among cognitive, behavioral, and resting brain states in data that are high dimensional, nonstationary, and only indirectly related to the underlying neural process. This is a setting in which topological methods are especially attractive because the state of interest is inherently distributed and relational. The target of inference is not a single variable but the organization of activity across many regions, times, or latent coordinates.

Saggar and colleagues used topological data analysis to reveal the dynamical organization of whole-brain activity maps at the single-participant level without arbitrarily collapsing data in space or time \cite{Saggar2018a}. Their Mapper-based framework tracked both within-task and between-task transitions and showed that the resulting topological representation was behaviorally informative. In later work, Saggar and colleagues applied a related approach to resting-state data and reported a highly visited transition state that behaves as a hub-like switch between neural configurations \cite{Saggar2022a}. These studies are especially important for the broader argument of this review because they treat topology not merely as a classifier of brain states, but as a means of representing the transition structure among those states.

Chung and collaborators extended this line in two complementary directions. In one study, they proposed a dynamic-TDA framework for functional brain networks that explicitly builds persistent homology over time-varying networks \cite{Chung2024a}. In another, they developed a topological state-space estimation framework for functional human brain networks, using topological distances to estimate and compare transient states of dynamic network organization \cite{Chung2024b}. Together, these papers help move topological neuroscience from static summaries of connectomes toward explicit modeling of state evolution and transition structure.

What makes these applications especially relevant to the theme of this review is that topology is being used to operationalize state reconfiguration itself. The system is the brain as a dynamic multivariate process; the construction is a dynamic activity cloud, Mapper graph, or time-indexed network; the topological observable is a state landscape, persistence summary, or topological distance; and the signal is a state-transition structure rather than only a scalar anomaly score. Compared with standard dynamic functional-connectivity pipelines, snapshot graph summaries, or cluster-based state segmentation, topological approaches preserve the geometry and bottlenecks of state-space organization. They therefore make it possible to ask not only which states exist, but how the system moves among them.

The limitations are familiar but important. The meaning of a loop, branch, or transition hub remains context dependent; preprocessing and distance choices matter substantially; and a topological transition state is not yet a neural mechanism. Even so, the neuroscience literature is among the clearest demonstrations that topology can turn a changing state landscape into an empirically useful signal. That is precisely what one would expect if topology is genuinely functioning as a language of emergent organization.

\subsection{Biological, ecological, and soft-matter transitions}

Biological and ecological systems broaden the evidentiary base by showing that topology can be useful even when the state variable of interest is diffuse, spatially extended, or only partially observed. In these domains, the warning problem often concerns vulnerability to undesirable state change rather than a sharply defined bifurcation point. This is important because it shows that topological warning does not require a textbook nonlinear-dynamics setting to be meaningful.

Syed Musa and colleagues provide a particularly clear example in hydrological early warning. They used persistent homology as a preprocessing step for flood early-warning signals and reported that the resulting topological signal, when coupled with critical-slowing-down indicators, produced fewer false alarms than an approach based directly on the raw water-level series \cite{Musa2021a}. This study is methodologically important because it makes the role of topology very explicit: persistent homology does not replace the warning system, but extracts a structurally informative signal that improves the performance of an otherwise familiar early-warning pipeline.

Larson and colleagues applied a broader set of TDA tools to decades of water-quality data from the Upper Mississippi River System in order to identify ecosystem states, characterize the variables associated with those states, and detect state transitions over time \cite{Larson2023a}. Their work is especially relevant because it reframes warning as monitoring \emph{vulnerability} to undesirable transitions rather than predicting a singular event. In this context, topology functions less as a crisis alarm and more as a state-space cartography tool for a large multivariate ecosystem. That is still very much within the logic of early warning, because management often depends on knowing when the system is drifting toward less desirable regimes.

The same logic appears in physical and soft-matter systems, where topological summaries are increasingly used as order-parameter surrogates for phase transitions and structural reorganization. Donato and colleagues used persistent homology to analyze phase transitions in canonical lattice models \cite{Donato2016a}. Tran, Chen, and Hasegawa developed a topological persistence machine for phase transitions, illustrating how persistence-based features can distinguish phases and critical regimes \cite{Tran2021a}. Membrillo Solis and colleagues then showed that topological structural heterogeneity can track the time evolution of soft-matter systems and detect order--disorder transitions from mesoscale organization \cite{MembrilloSolis2022a}. In a more explicitly biological direction, Spector, Harrington, and Gaffney used persistent homology to classify parameter-dependent pattern changes in Turing reaction--diffusion systems, suggesting that topology can act as a state descriptor for self-organizing biological pattern formation \cite{Spector2026a}.

These studies collectively suggest that topology is often most valuable in biological and ecological settings when no obvious low-dimensional order parameter exists. The system may be heterogeneous, spatially distributed, or morphologically complex; the construction may involve images, spatial point patterns, sublevel sets, or multivariate environmental states; and the topological observable functions as a descriptor of mesoscale organization. In such situations, topology can serve as a practical surrogate for ``state'' itself. The caution, however, is that genuinely prospective warning studies remain fewer here than in finance or hydrology, and many results are retrospective or simulation based. Topology is clearly useful for detecting and classifying transitions in these domains; the case for robust ex ante forecasting is promising but still less mature.

\subsection{Engineered and networked systems}

Engineered systems make the operational demands on topological methods especially clear. In cybersecurity, industrial monitoring, infrastructure telemetry, and network traffic analysis, an anomaly is often not a large excursion of one variable but a reorganization of event structure or dependence across many channels. Pointwise thresholds and local features can miss such changes precisely because the anomaly is global or structural. This is a domain in which topological methods are naturally aligned with the problem.

Bruillard, Nowak, and Purvine provided an early demonstration in cybersecurity by applying persistent homology to computer logs and showing that topological and spectral embeddings contained discriminative information complementary to standard count-based log embeddings \cite{Bruillard2016a}. Their result is conceptually important because it illustrates a recurring theme of this review: topology is often most useful when it supplements, rather than replaces, conventional feature engineering. The gain comes from preserving global structure that ordinary summary vectors ignore.

More recent work has moved from proof-of-concept toward general anomaly-detection pipelines. Chazal, Levrard, and Royer introduced TADA, a method for detecting structural anomalies in dependent sequences such as multivariate time series by monitoring topological changes in dynamic dependence graphs \cite{Chazal2024a}. Their numerical results suggest that topological embeddings are especially suited to detecting global changes in correlation structure rather than only local point anomalies. Moitra, Malott, and Wilsey addressed a different bottleneck by developing a streaming framework for persistent homology via topological data summaries, including an application to anomalous network-traffic bursts \cite{Moitra2023a}. Dee Algar and colleagues' regime-change work on time series also fits naturally into this operational cluster because it treats topological distances as online indicators of behavioral change rather than as static descriptive summaries \cite{DeeAlgar2021a}.

The pattern here is strikingly consistent with the rest of the section. The system is a networked or multichannel engineered process; the construction is a dynamic graph, log-derived complex, or streaming summary; the topological observable is a persistence-based embedding or distance; and the signal is an anomaly score or change statistic. Topology is strongest when the abnormality manifests as a structural departure from a normal regime rather than as a single extreme value. This is also the setting in which the limitations of topology become most operationally visible. Real-time deployment requires streaming computation, threshold calibration, and false-positive control. Topological observables therefore matter only insofar as they can be integrated into a larger monitoring architecture.

Compared with finance or neuroscience, the literature on autonomous swarms and broader socio-technical infrastructures remains thinner. That is less a weakness of the present review than an indication of where the next wave of applications may emerge. The conceptual fit is clear: these systems are multiscale, nonlinear, and structurally coordinated, and their failures are often collective reorganizations rather than isolated local events. What remains underdeveloped is the transition from mathematically suggestive topology to deployable warning systems in these domains.

Taken together, these applications support a common conclusion. Topology contributes most when anomalies or transitions are reorganizations of collective structure and when the resulting topological object can be converted into a monitored statistic. The specific constructions vary widely---delay embeddings, correlation networks, functional connectomes, ecological state spaces, streaming dependence graphs---but the recurring logic is the same. Topology becomes practical when it turns reorganizing structure into measurable signal. This is why its empirical success in detection and early warning is not accidental. It follows directly from the representational role developed in the previous sections. The next section builds on this point by asking when, across these domains, topology is most likely to work well and when its apparent advantages begin to break down.

Table~\ref{tab:applications-summary} condenses the applications literature reviewed in this section into a common template: how the system is rendered, what topological construction is used, what observable is monitored, and what competing baseline it is meant to complement or outperform.

\begin{table}[t]
\centering
\small
\setlength{\tabcolsep}{4pt}
\renewcommand{\arraystretch}{1.15}
\begin{tabularx}{\textwidth}{L{0.13\textwidth} L{0.17\textwidth} L{0.16\textwidth} L{0.13\textwidth} Y Y}
\toprule
\textbf{Domain} &
\textbf{Typical rendering} &
\textbf{Topological construction} &
\textbf{Typical observable} &
\textbf{What topology contributes} &
\textbf{Primary comparison target} \\
\midrule

Nonlinear dynamics &
Sliding windows of state trajectories or delay embeddings &
Vietoris--Rips / \v{C}ech filtrations; sometimes sublevel or cubical constructions &
Betti curves, persistence landscapes, diagram distances &
Tracks attractor reorganization, bifurcation-related structural change, and regime shifts &
Classical early-warning indicators, bifurcation analysis, recurrence analysis, manifold methods \\

Finance and market stress &
Time-varying correlation or dependence networks; embedded price series &
Clique/flag filtrations or point-cloud persistence &
Persistence landscapes, entropy, Wasserstein or bottleneck distances &
Converts market-wide reorganization into structural stress signals &
Volatility, correlation spikes, PCA, clustering, factor-style summaries \\

Neuroscience and brain-state transitions &
Functional-connectivity networks, co-activation data, state-space clouds &
Clique/concurrence complexes, Mapper, homological scaffolds &
Cavities, branches, state transitions, scaffold summaries &
Reveals mesoscale coordination and state organization not captured by edge summaries alone &
Dynamic-connectivity baselines, graph metrics, state-space clustering \\

Biology, ecology, and soft matter &
Morphospaces, ecological state vectors, material or configuration data &
Geometric filtrations and higher-order constructions adapted to the domain &
Persistence diagrams, loops/cavities, phase- or morphology-related summaries &
Identifies reorganizing structure in morphology, ecosystems, and material order &
Domain order parameters, geometric descriptors, mechanistic models \\

Engineered and networked systems &
Logs, sensor streams, dynamic graphs, multichannel monitoring data &
Streaming filtrations, graph-to-complex lifts, simplicial signal spaces &
Topological distances, anomaly scores, Hodge spectra &
Turns structural departure from baseline into monitored change statistics &
Change-point detection, graph anomaly detection, forecasting residuals, control-theoretic monitors \\

\bottomrule
\end{tabularx}
\caption{Applications-by-domain summary. Across the domains reviewed in Section~7, the recurring pattern is that topology becomes useful when reorganizing structure can be rendered as a monitored statistic.}
\label{tab:applications-summary}
\end{table}

\section{Cross-Domain Synthesis: When Does Topology Work Best?}
\label{sec:cross_domain_synthesis}

By this point in the review, the literature is broad enough to support a comparative judgment rather than a simple catalog of applications. Topological methods have now been used across nonlinear dynamics, finance, neuroscience, ecology, materials science, and engineered monitoring systems, and the pattern that emerges is neither that topology is universally superior nor that it is merely a fashionable add-on \cite{Wasserman2018a,Chazal2021a,Gidea2018a,Saggar2018a,Larson2023a,Chazal2024a}. Instead, topology appears to perform best under a relatively specific set of conditions. It is most effective when the relevant signal is \emph{organizational}, when there is no obviously correct single scale or threshold, when the representation preserves latent geometry or higher-order relations, and when topology is used either as a structural diagnostic or as a feature generator inside a broader analytic pipeline. It is less effective when the data are too sparse, when the chosen metric or filtration distorts the intrinsic structure, or when the scientific target is dominated by non-topological covariates.

Table~\ref{tab:question-matrix} restates the cross-domain synthesis as a practical decision matrix, making explicit which kinds of scientific questions call for which representational choices, what topological objects and observables they motivate, which non-topological baselines remain strong competitors, and where the most common failure modes arise.

\begingroup
\small
\setlength{\tabcolsep}{4pt}
\renewcommand{\arraystretch}{1.15}

\begin{longtable}{L{0.14\textwidth} L{0.15\textwidth} L{0.15\textwidth} L{0.14\textwidth} L{0.18\textwidth} L{0.18\textwidth}}
\caption{Question-to-method matrix for topological analysis in complex systems.}
\label{tab:question-matrix}\\
\toprule
\textbf{Scientific question} &
\textbf{Representation choice} &
\textbf{Topological object / procedure} &
\textbf{Observable} &
\textbf{Strong competitors or baselines} &
\textbf{Common failure modes} \\
\midrule
\endfirsthead

\toprule
\textbf{Scientific question} &
\textbf{Representation choice} &
\textbf{Topological object / procedure} &
\textbf{Observable} &
\textbf{Strong competitors or baselines} &
\textbf{Common failure modes} \\
\midrule
\endhead

\midrule
\multicolumn{6}{r}{\small\itshape Continued on next page}
\endfoot

\bottomrule
\endlastfoot

Is a dynamical system approaching a regime shift? &
Delay embedding or sliding window state cloud &
Vietoris--Rips / \v{C}ech filtration on successive windows &
Persistence distance, landscape change, Betti trajectory &
Variance/autocorrelation early-warning signals, recurrence analysis, manifold learning &
Poor embedding, bad window size, nonstationarity, sparse sampling, wrong metric \\

Is weighted relational structure entering a stressed regime? &
Time-varying correlation or similarity network &
Clique or flag filtration; weighted-network persistence &
Topological distance, entropy, Betti curves over time &
Volatility, eigenvalue spectra, community change, clustering &
Spurious dependencies, preprocessing sensitivity, threshold artifacts, domination by common mode \\

Are higher-order closure or group interactions scientifically decisive? &
Observed co-activity, co-participation, or multiway event data &
Simplicial complex or hypergraph, possibly with Hodge operators &
Cavities, scaffold structure, harmonic modes &
Graph measures, motifs, pairwise contagion or synchronization models &
Inferred rather than observed higher-order structure, unjustified closure, sparse participation \\

Is coarse state-space branching more important than invariant hole counting? &
Feature space equipped with a domain-relevant lens and local clustering rule &
Mapper, multiscale Mapper, or Reeb-type summary &
Branch structure, bottlenecks, coarse state transitions &
Clustering, PCA/UMAP trajectories, hidden Markov models &
Lens sensitivity, cover and overlap dependence, clustering instability, over-reading visual summaries \\

Is online anomaly detection or operational monitoring the goal? &
Streaming windows on graphs, complexes, or multichannel signals &
Repeated persistence, streaming filtrations, or operator-based monitoring &
Change statistic, anomaly score, Hodge spectral shift &
Change-point detection, graph anomaly detectors, forecasting residuals &
Computation, calibration, concept drift, false positives, difficult integration into existing pipelines \\

Do alternate renderings of the same system materially change the conclusion? &
Parallel candidate renderings of the same data &
Pipeline-level comparison across multiple topological constructions &
Performance, interpretability, and stability across renderings &
Domain-informed ablations and strongest existing baseline pipeline &
Mistaking preprocessing gains for topological gains; weak causal attribution of performance \\

\end{longtable}
\endgroup

The goal of this section is therefore evaluative. Rather than asking whether topology \emph{can} be used in a domain, the more useful question is when it is likely to add value beyond standard statistical, geometric, or graph-based descriptions. The answer, across the literature reviewed so far, is surprisingly coherent.

\subsection{High-dimensionality helps only when the structure is recoverable}

A recurring temptation in the applied literature is to say that topology is especially appropriate for high-dimensional data. That claim is true only in a qualified sense. Topology is useful not because data live in a high-dimensional ambient space, but because such data often contain \emph{latent} geometric or relational organization that is difficult to summarize with a small number of scalar descriptors. Delay embeddings of periodic signals, neural activity clouds, multivariate ecosystem states, and market-dependence structures all fit this pattern: they are high dimensional in presentation, but constrained by lower-dimensional or otherwise structured organization in the relations among observations \cite{Perea2015a,Saggar2018a,Gidea2018a,Larson2023a}. In such settings, topology can recover robust global features---loops, branches, cavities, transition bottlenecks---that are obscured by local coordinate variation.

Turke\v{s}, Mont\'ufar, and Otter make this point especially clearly in benchmark form. Their analysis suggests that persistent homology is most effective when the classes or regimes to be distinguished differ in topological or closely related geometric properties such as the number of holes, curvature, or convexity \cite{Turkes2022a}. Importantly, they also show that persistent-homology features can remain useful under limited training data, noisy test data, and restricted computational budgets when the target really is structural \cite{Turkes2022a}. That finding helps explain why TDA has been attractive in domains where high-dimensional observations are abundant but labels, mechanistic certainty, or clean order parameters are scarce.

At the same time, recent work makes clear that high-dimensionality alone is not enough. Damrich, Berens, and Kobak show that standard persistent homology based on Euclidean distances can become extremely sensitive to noise when the data have low intrinsic dimensionality but are embedded in a much higher-dimensional ambient space \cite{Damrich2024a}. Their examples are instructive precisely because the underlying topology is simple: even a noisy circle in \(\mathbb{R}^{50}\) can defeat a naive Euclidean pipeline \cite{Damrich2024a}. The lesson is not that topology fails on high-dimensional data, but that it succeeds only when the representation preserves intrinsic geometry rather than ambient noise. In practice, this means that metric choice, graph construction, denoising, dimensionality reduction, or alternative filtration design are often decisive.

This point can be stated more generally. The favorable regime for topology is \emph{high-dimensional but structured} data, not high-dimensional data in the abstract. When the latent organization is recoverable, topology can expose it. When the representation is dominated by irrelevant degrees of freedom, topology can be misled just as readily as any other method.

\subsection{Topology contributes most when the target is organizational rather than scalar}

The applied literature also suggests a second, more important regularity. Topology adds the most when the scientific question is about \emph{organization} rather than about a small number of scalar parameters. This is why topological methods recur in studies of regime shift, anomaly detection, state reconfiguration, and higher-order coordination. In all of these cases, the target of inference is not simply a change in magnitude but a change in how the system is arranged.

The contrast with classical early-warning signals makes this particularly clear. Scheffer and colleagues established the modern early-warning framework around generic scalar indicators such as rising variance and autocorrelation near critical transitions \cite{Scheffer2009a}. Those indicators remain important, but many of the topological studies reviewed above target a different layer of description. Gidea and Katz, for example, use persistence-based summaries to study market-wide reorganization around crisis periods; Saggar and colleagues use Mapper-like constructions to map the changing organization of whole-brain states; and Chazal, Levrard, and Royer detect anomalies in dependent sequences by monitoring structural changes in dynamic dependence graphs \cite{Gidea2018a,Saggar2018a,Saggar2022a,Chazal2024a}. In each case, topology is useful because the phenomenon of interest is a reconfiguration of relations before it is a dramatic excursion in one or two variables.

This is also why topology often performs well when there is no clear parametric model or low-dimensional order parameter. In nonlinear dynamics, the changing shape of a reconstructed attractor can be more informative than a single summary statistic \cite{Mittal2017a,DeeAlgar2021a}. In ecology, topological state-space methods can describe vulnerability to regime change even when the ecosystem cannot be reduced to one canonical control variable \cite{Larson2023a}. In neuroscience, the transition structure among distributed brain states is often more informative than any one edge weight or regional activation level \cite{Saggar2018a,Chung2024b}. Topology is well matched to such problems because it does not require one to know in advance which scalar variable should carry the signal. It allows the signal to reside in the \emph{organization} of the representation itself.

A closely related pattern appears in higher-order systems. When pairwise graphs are too coarse to preserve the mechanism of interest, topological or topology-inspired formalisms gain a comparative advantage. Battiston and colleagues, Bick and colleagues, and Zhang and collaborators all show that hypergraphs and simplicial complexes can preserve many-body structure and closure relations that ordinary graphs erase, with direct consequences for contagion, synchronization, and collective dynamics \cite{Battiston2020a,Bick2023a,Zhang2023a}. In such cases, topology is not merely a \textit{post hoc }summary. It is part of the representational language required to make the right object visible in the first place.

\subsection{Topology is especially valuable when threshold ambiguity is part of the problem}

Another recurring advantage of topology is less glamorous but methodologically important: it handles threshold ambiguity better than many competing representations. Weighted networks, similarity graphs, correlation matrices, and distance clouds almost never come with one naturally privileged cutoff. Yet a great deal of downstream analysis in complex systems depends on committing to such a cutoff before interpretation begins. Persistent homology and related multiscale constructions are attractive precisely because they replace a single threshold with a family of thresholds and then ask which structures survive across them \cite{Carlsson2009a,Edelsbrunner2008a}.

This feature matters in practice. Garrison and colleagues show that standard graph-theoretic measures of functional brain networks can vary substantially across thresholds, sometimes even altering the qualitative interpretation of group differences \cite{Garrison2015a}. Similar concerns arise in finance when correlation networks are thresholded, in ecology when similarity graphs are sparsified, and in anomaly detection when dependence structures are binarized. Topology does not remove these modeling choices entirely, but it changes their role. Instead of hiding the threshold inside preprocessing, it turns threshold variation into part of the observable itself.

This is one reason persistence-based methods are so prominent in the transition-detection literature. Ismail and colleagues' financial-crisis studies, for example, depend explicitly on how market structure changes across correlation constructions and filtration levels \cite{Ismail2022a,Ismail2022b}. In nonlinear dynamics, CROCKER plots, persistence landscapes, and diagram trajectories likewise monitor structural change across scale rather than at one arbitrarily fixed resolution \cite{Guzel2022a,DeeAlgar2021a}. The common insight is that topology becomes especially compelling when the analyst's uncertainty about scale is not a nuisance to be suppressed, but part of the scientific problem.

This point also helps explain why topology is often more persuasive as a \emph{diagnostic} than as a black-box predictor. If the main difficulty is that the system may reorganize in different ways at different resolutions, then a multiscale topological summary is informative even before it is converted into a predictive score. That diagnostic role is easy to overlook when persistent homology is discussed only as a feature-engineering tool, but it is central to its value in complex-systems work.

\subsection{Across domains, topology is strongest as a structural diagnostic or feature extractor}

The application literature reviewed so far supports a further synthesis: topological methods are usually at their best when they operate either as structural diagnostics or as feature generators for downstream methods, rather than as fully autonomous predictive models. This is not a weakness. It is a reflection of what topology actually contributes.

In many successful applications, the first step is not a direct prediction but the construction of a new state variable. Persistence landscapes, images, Betti curves, diagram distances, and Mapper-derived transition graphs all convert distributed organization into objects that can be averaged, clustered, thresholded, or fed into machine-learning models \cite{Bubenik2015a,Adams2017a,Saggar2018a,Chazal2024a}. This pattern is visible in regime-change detection, where topological summaries become warning statistics; in neuroscience, where they become state-space maps; and in anomaly detection, where they become structural deviation scores \cite{Myers2019a,Saggar2022a,Chazal2024a}. In that sense, topology is rarely the whole pipeline. It is the layer that makes organization measurable.

The comparative evidence from learning tasks points in the same direction. Di Via, Di Via, and Fugacci show that persistent-homology-based methods can outperform standard machine-learning models in extremely low-label settings, while hybrid PH+ML approaches often improve as the amount of labeled data increases \cite{DiVia2024a}. Their results are striking not because they prove the universal superiority of topology, but because they illustrate a recurring tradeoff. Purely topological methods can be remarkably data efficient when class differences are genuinely structural, whereas hybrid methods often become preferable once enough data exist to exploit complementary non-topological cues \cite{DiVia2024a}. This is closely aligned with the broader benchmark results of Turke\v{s}, Mont\'ufar, and Otter, who likewise emphasize that persistent homology is most effective when the target phenomenon is structural and sufficiently aligned with the inductive bias of the representation \cite{Turkes2022a}.

The resulting empirical picture is consistent across domains. Topology is strongest as a \emph{structural diagnostic}, a \emph{state descriptor}, or a \emph{feature extractor}. It is sometimes competitive as a standalone classifier or detector, especially in data-scarce or highly structured settings, but its most robust role is often to provide observables that other inferential frameworks can then exploit. That is exactly what one should expect if topology is primarily a language of organization rather than a universal predictive engine.

\subsection{Failure modes and boundary conditions}

The comparative success of topology becomes easier to interpret once its failure modes are stated plainly. The first boundary condition is sampling. Niyogi, Smale, and Weinberger's homology-inference results make clear that topological recovery depends on sufficient sampling density and geometric regularity \cite{Niyogi2008a}. That observation remains important for modern applications. If the data are too sparse relative to the scale of the feature one hopes to recover, then the inferred topology may reflect the sample more than the underlying system. No amount of sophistication in vectorizing persistence diagrams can fully compensate for a representation that never captured the relevant structure in the first place.

The second boundary condition is metric and filtration mismatch. Damrich, Berens, and Kobak show one contemporary version of this problem in high-dimensional noisy data, where Euclidean distances obscure intrinsic loops that become visible only after a better distance construction is chosen \cite{Damrich2024a}. Garrison and colleagues show a related issue from another angle: threshold choices in weighted networks can materially alter downstream structure \cite{Garrison2015a}. Torres and colleagues' general lesson about representation therefore applies with full force here. Topology does not eliminate modeling judgment. It relocates that judgment to the choice of metric, filtration, cover, complex, or relational object \cite{Torres2021a}.

A third limitation is semantic thinness. Topological features are intentionally abstract, and that abstraction is part of their power. But it also means that a loop, cavity, or branch is not self-interpreting. Its scientific meaning depends on the representation from which it was computed and on the domain-specific process that generated that representation. Recent review work by Su and colleagues underscores this point by noting that persistent homology can be limited by its high-level abstraction and by its insensitivity to changes that are important scientifically but not topological in character \cite{Su2026a}. In some problems, that abstraction is exactly what makes topology useful. In others, it can cause topological summaries to ignore variables that are decisive for prediction or explanation.

This leads to a final and perhaps simplest failure mode: topology is weak when the signal of interest is not predominantly topological. If the scientifically relevant distinction lies in local intensity differences, effect sizes, signed directionality, or variables that do not appreciably reorganize the qualitative structure of the representation, then topological methods may add little. This is not a criticism of topology so much as a reminder that every representation carries an inductive bias. Persistent homology privileges robustness of connectedness, loops, and cavities across scale. It does not promise sensitivity to every kind of change that may matter in a complex system.

\subsection{Comparative conclusion}

Taken together, the cross-domain literature suggests a fairly crisp synthesis. Topology works best when four conditions are met. First, the phenomenon of interest is distributed and organizational rather than reducible to a few local or scalar variables. Second, there is meaningful structure across scale, threshold, or interaction order, so that a multiscale or higher-order representation is scientifically justified. Third, the chosen metric or relational construction respects the intrinsic structure of the data rather than distorting it with ambient noise or arbitrary thresholding. Fourth, topology is allowed to function in the role it serves best: as a structural diagnostic, a state descriptor, or a feature generator that can be integrated with statistical, dynamical, or machine-learning methods.

Conversely, topology struggles when any of these conditions fail. Sparse sampling, poor filtrations, unstable thresholding, high-dimensional noise under inappropriate metrics, and tasks dominated by non-topological covariates can all undermine its apparent advantages \cite{Niyogi2008a,Garrison2015a,Damrich2024a,Su2026a}. The point is not that these are merely practical inconveniences. They define the boundary of the representational regime in which topology is genuinely informative.

This conclusion helps sharpen the overall thesis of the review. The empirical success of topological methods in complex systems does not come from a mysterious general power of ``shape.'' It comes from a much more specific fit between topological representations and scientific problems that are about multiscale, higher-order, and emergent organization. That fit is strong, but it is not unlimited. The next section therefore turns from comparative synthesis to a more explicit discussion of the methodological limitations and open problems that currently shape the field.

\section{Limitations and Open Problems}
\label{sec:limitations_open_problems}

The case made so far for topological methods in complex-systems science is strong, but it is not unconditional. In fact, many of the field's most significant limitations arise from the same features that give topology its value. Because topological methods are designed to abstract away from local detail and preserve qualitative organization, they inevitably force difficult choices about representation, significance, interpretation, and scale. Those difficulties are not peripheral annoyances. They define the present frontier of the subject.

This is especially important for the kind of review advanced here. If topology is to be taken seriously as a language for emergent organization, then it cannot be treated as a purely decorative add-on to existing pipelines. It must instead be judged by the same standards applied to any other scientific representation: Does it isolate the right object? Under what assumptions is the resulting summary statistically meaningful? What exactly do its observables tell us about mechanism, structure, or prediction? And when do computational and methodological costs outweigh the benefits? The contemporary literature has made real progress on each of these questions, but it has not resolved them \cite{Wasserman2018a,Chazal2021a,Otter2017a,Su2026a}.

The purpose of this section is therefore twofold. First, it identifies the main limitations that currently constrain the use of topological methods in complex-systems research. Second, it recasts those limitations as open problems that define a forward research agenda. The overall claim is not that topology is too immature to be useful. It is that the field is mature enough for its most difficult unresolved questions to become visible.

\subsection{Representation dependence is irreducible}

A first limitation is also the most fundamental: topological output is inseparable from representational choice. Persistent homology, Mapper, and higher-order constructions do not operate on raw reality. They operate on point clouds, weighted graphs, simplicial complexes, delay embeddings, cover constructions, or other mathematically mediated objects. Torres and colleagues emphasize that this is true of all scientific representations, not only of topology, but the point bears repeating here because topological summaries can easily appear more objective than they really are \cite{Torres2021a}. A persistence diagram may look like a direct description of structure, but it is always the description of structure \emph{relative to} a chosen metric, filtration, embedding, or complex construction \cite{Chazal2021a,Wasserman2018a}.

This dependence matters particularly strongly in complex systems because the relevant notion of proximity or relation is seldom unique. In a financial application, the analyst may build a filtration from Pearson correlations, rank correlations, delay embeddings, or some learned similarity. In neuroscience, the object might be a co-activation complex, a thresholded functional-connectivity matrix, a dynamic point cloud, or a latent representation. In ecology, one might use environmental-state vectors, spatial proximity, species co-occurrence, or transition trajectories. None of these choices is simply technical. Each is a hypothesis about where the system's organization resides.

Stability theorems do not remove this dependency. The stability results of Cohen-Steiner, Edelsbrunner, and Harer are powerful precisely because they show that persistence diagrams are robust under perturbations of a \emph{fixed} filtration function \cite{CohenSteiner2007a}. But they do not determine which filtration should have been chosen in the first place. A stable answer to the wrong representational question can still be scientifically misleading. The interpretive burden therefore shifts upstream: one must justify the object on which topology is computed, not only the topology extracted from it.

Recent work by Damrich, Berens, and Kobak provides a sharp demonstration of this point. They show that persistent homology based on naive Euclidean distances can fail dramatically for data with low intrinsic dimensionality embedded in a high-dimensional noisy ambient space, whereas spectral distances on \(k\)-nearest-neighbor graphs can recover the correct topology much more reliably \cite{Damrich2024a}. Their result is instructive because it does not merely identify a practical nuisance. It shows that the question of the data topology is often ill posed until a suitable metric or relational geometry has been specified. In complex-systems applications, where ambient noise, hidden variables, and heterogeneous observables are common, this issue is likely to be the rule rather than the exception.

This representational problem becomes even more acute when one moves beyond one-parameter persistence. Many complex systems are organized simultaneously by several notions of scale: spatial resolution, temporal aggregation, density, interaction strength, directionality, or uncertainty. Compressing all of that into a single filtration parameter can be expedient, but it can also be conceptually distorting. Botnan and Lesnick argue that multiparameter persistence is a natural response to this difficulty, while also emphasizing that defining barcodes in the multiparameter setting is problematic and that practical applications remain comparatively immature \cite{Botnan2023a}. Carlsson and de Silva's zigzag persistence addresses a related difficulty by allowing one to study families of spaces connected by both forward and backward maps, thereby handling situations not covered by ordinary monotone filtrations \cite{Carlsson2010a}. These developments are promising, but they do not abolish representational judgment. They make it richer, and in some respects harder.

The upshot is simple: there is no topology of a complex system independent of the way the system has been rendered topological. This is not a defect unique to TDA, but it is one of the main reasons the method can fail when applied without domain-specific care. This limitation should be made fully explicit. Topology is powerful precisely because it preserves organization under a chosen representation. It does not decide that representation for the analyst.

\subsection{Stability is not significance: inference and null models}

A second major limitation concerns statistical meaning. The existence of a persistent feature does not by itself establish that the feature is scientifically significant. This distinction is easy to blur because the language of persistence invites a natural but incomplete heuristic: long-lived features are signal; short-lived features are noise. That heuristic is often useful, but it is not a substitute for inference.

Important progress has been made on this problem. Fasy and colleagues developed confidence sets for persistence diagrams, while Chazal and collaborators developed robust topological inference procedures based on distance-to-a-measure and related constructions \cite{Fasy2014a,Chazal2018a}. Wasserman's review is also valuable here because it frames TDA explicitly as a statistical enterprise rather than as a purely geometric curiosity \cite{Wasserman2018a}. These contributions show that persistent topological summaries can be treated as inferential objects. But they do not imply that inference from topological summaries is generically straightforward.

Recent work by Vishwanath, Fukumizu, Kuriki, and Sriperumbudur makes this point especially clearly. They study necessary and sufficient conditions under which valid statistical inference is possible using topological summary statistics, and they give examples of models that are invariant with respect to such summaries \cite{Vishwanath2025a}. The importance of this result is conceptual as much as technical. It means that there are settings in which no amount of clever downstream statistics can recover distinctions that the chosen topological summary has erased. Inference can fail not only because sample sizes are small or models are misspecified, but because the summary itself is insufficiently informative for the statistical question being asked.

This limitation is closely tied to the problem of null models. In practical complex-systems work, one rarely wants to know whether a persistence diagram is surprising relative to complete randomness. One wants to know whether it is surprising relative to a specific family of constrained alternatives: random geometric noise with the same density, a correlation structure with the same marginal distributions, a network with the same degree or strength sequence, a spatially embedded null with the same wiring-cost profile, or a time series with the same autocorrelation and seasonality. Bobrowski and Skraba's universal null-distribution for persistence diagrams is a substantial advance because it provides a new hypothesis-testing framework for point-cloud settings \cite{Bobrowski2023a}. Yet many complex-systems applications do not live naturally in that regime.

This is where domain-specific null-model work becomes indispensable. Váša and Mišić's review of null models in network neuroscience is especially instructive, not because it is itself a TDA paper, but because it shows how carefully a scientifically meaningful null must preserve or randomize different structural properties depending on the question at hand \cite{Vasa2022a}. The same lesson applies more broadly. A topological feature in a correlation-derived brain network, an ecological state-space reconstruction, or a financial dependence graph is only as meaningful as the null hypothesis against which it has been tested. Generic random-point-cloud baselines are often too weak. They can tell us that a structure is not trivial, but not whether it is surprising once the obvious non-topological constraints of the system have been respected.

For complex-systems science, this is a central open problem. The field needs null models that are not merely mathematically convenient, but scientifically matched to weighted, temporal, correlation-derived, higher-order, and spatially constrained data. Without that, topology risks becoming strongest where the null is weakest. That would be a poor foundation for a rigorous review or for robust downstream science.

\subsection{Interpretability remains a genuine bottleneck}

A third limitation concerns interpretability. Persistent homology is often praised for being more interpretable than many machine-learning features because connected components, loops, and cavities have intuitive meanings. That praise is justified only up to a point. The difficulty is that persistence diagrams describe the \emph{existence} and lifetime of homological features, but usually not their unique realization in the underlying data. In other words, they tell us that some robust organization exists, but not always which concrete geometric or relational structure is responsible for it.

This ambiguity is not a minor technicality. Li and colleagues show that cycle representatives of persistent classes are highly non-unique and that different choices of representative can lead to quite different interpretations of the same birth--death pair \cite{Li2021a}. Their study is valuable because it makes explicit a tension that many applications only encounter implicitly. If one wants to map a persistent 1-cycle in a brain network, a molecular configuration, or a time-delay embedding back to a concrete scientific object, there may be many mathematically valid ways to do so. The diagram itself does not choose among them.

Recent work has begun to address this issue. Obayashi's stable volumes are designed to identify more robust geometric structures associated with persistence pairs and to reduce the noise sensitivity of earlier optimal-volume constructions \cite{Obayashi2023a}. This is an important step because it moves the field from merely computing persistence summaries toward locating stable representatives in the data. Even so, the broader interpretive problem remains. A representative cycle or volume is still only a candidate explanation of a topological feature, and its scientific meaning depends on the domain-specific semantics of the underlying space.

This matters acutely in complex-systems applications because the analyst often wants more than detection. One wants to know \emph{what reorganized}. If a persistence landscape rises before a market shock, which sectoral or asset-level configuration is responsible? If a cavity appears in a connectome analysis, what kind of functional or anatomical constraint does it reflect? If a topological anomaly is detected in a cyber-physical system, which channels or subsystems are implicated? These are not questions about topology alone. They are questions about the mapping from topological summaries back to mechanistic or semantic structure.

For that reason, interpretability in TDA should not be understood merely as human readability of diagrams. The real open problem is \emph{semantic grounding}: connecting a topological feature to a stable, domain-legible structure in the underlying system. Representative-cycle methods, stable-volume constructions, cohomological coordinate systems, and related tools help, but they do not yet provide a general solution. In the context of this review, that limitation should be emphasized rather than hidden. If topology is to function as a language for emergent organization, it must do more than detect structure; it must increasingly help identify which emergent structure is present and where it lives.

\subsection{Computation and scale remain constraining factors}

A fourth limitation is computational. Persistent homology has benefited from major algorithmic advances, but the cost of computing topological summaries at realistic scales remains substantial, especially in the very applications where complex-systems researchers are most interested: high-dimensional data, repeated sliding windows, large weighted networks, streaming inputs, and higher-order constructions.

Otter and colleagues' roadmap already made clear that the computational pipeline for persistent homology is highly sensitive to filtration type, software implementation, and data representation \cite{Otter2017a}. Bauer's Ripser paper marked a major advance for Vietoris--Rips persistence by combining cohomological computation, implicit boundary representation, and related optimizations to achieve substantial gains in speed and memory efficiency \cite{Bauer2021a}. Malott, Chen, and Wilsey then reviewed the broader high-performance-computing landscape and emphasized that large-scale persistent homology remains computationally expensive and memory intensive despite these advances \cite{Malott2023a}. The practical lesson is that TDA has become dramatically more feasible, but not cheap.

This issue is especially acute in complex-systems applications because topology is often applied not once but repeatedly. Sliding-window analyses in finance, neuroscience, anomaly detection, and nonlinear dynamics require persistence to be computed across many overlapping windows. Early-warning pipelines may then add bootstrapping, null-model evaluation, or multiscale comparisons on top of that. The computational challenge is not only the size of a single filtration, but the cumulative cost of monitoring a system through time.

Interpretability and computation also interact in unfortunate ways. The most interpretable topological analyses are often not the cheapest ones. Recovering representative cycles, stable volumes, or matched features across windows can be significantly more demanding than computing the diagrams alone \cite{Li2021a,Obayashi2023a}. Likewise, the move from one-parameter to multiparameter or zigzag settings often introduces richer scientific structure at the cost of much more difficult computation and visualization \cite{Botnan2023a,Carlsson2010a}.

From the perspective of complex systems, this is not just an engineering issue. Real-time monitoring, online anomaly detection, and adaptive control all require topological summaries that are fast enough to be operational and stable enough to be trustworthy. The open problem is therefore twofold: how to reduce the computational cost of topology on large or streaming data, and how to do so without sacrificing the structural fidelity that makes topology scientifically valuable in the first place.

\subsection{The one-parameter paradigm is too narrow for many systems}

A fifth limitation is conceptual. Much of the success of persistent homology has been built on the one-parameter paradigm: one defines a single filtration parameter, tracks births and deaths, and summarizes the result with barcodes or persistence diagrams. This framework is mathematically elegant and practically effective, but many complex systems are not naturally organized by one parameter alone.

Time-varying systems provide the clearest example. A dynamic brain network, a changing market-dependence structure, or an adaptive multi-agent system often undergoes both addition and deletion of structure over time. Ordinary persistence can be applied to snapshots or sliding windows, but it does not directly express the fact that structure may appear, disappear, and reappear in ways that are not monotone. Zigzag persistence was developed precisely to study families of spaces with both forward and backward maps, thereby extending persistence to situations outside the standard filtration setting \cite{Carlsson2010a}. That extension is highly relevant to complex systems, yet it remains much less common in applications than the one-parameter toolkit.

The same issue arises when one wants to track several notions of scale simultaneously. A complex system may be organized at once by density, time, confidence, interaction strength, or spatial resolution. Botnan and Lesnick's introduction to multiparameter persistence makes clear both why this generalization is natural and why it remains difficult in practice \cite{Botnan2023a}. In the multiparameter setting there is no single barcode with all the pleasant completeness and visualization properties familiar from one-parameter persistence; practical invariants, metrics, computation, and software are still under active development \cite{Botnan2023a}. The resulting situation is intellectually rich but operationally challenging. Exactly the systems that most naturally call for richer topological formalisms are often those for which the supporting methodology is least mature.

This limitation also helps explain the recent push beyond persistent homology narrowly understood. As Su and colleagues argue, persistent homology can be limited by its high level of abstraction, its insensitivity to some non-topological changes, and its difficulty with richer data types or additional structure \cite{Su2026a}. This does not make persistent homology obsolete; it means that complex-systems science will likely need a broader topological toolkit, including multiparameter methods, zigzag persistence, harmonic and Hodge-theoretic constructions, persistent operators, and possibly sheaf-theoretic or directed frameworks, depending on the domain.

The open problem here is not simply to generalize for the sake of generality. It is to determine which topological formalism matches which kind of complex-system representation. One-parameter persistence is powerful because it is clear. The next generation of topological methods must become comparably usable without losing the extra structure that motivates them.

\subsection{Benchmarking and evaluation are still too uneven}

A final limitation concerns evaluation culture. Much of the literature on topological methods in complex systems remains organized around successful case studies. Those case studies are often insightful, but they do not always answer the comparative question that matters most for a mature review: when does topology outperform strong alternatives, under what conditions, and by which criteria?

There has been progress on this front. Turke\v{s}, Mont\'ufar, and Otter provide one of the clearest studies of when persistent homology is effective, and Di Via, Di Via, and Fugacci offer a useful comparison between persistent-homology-based methods and learning-based methods in limited-data settings \cite{Turkes2022a,DiVia2024a}. These papers are important because they treat TDA as a method whose strengths and weaknesses can be benchmarked rather than assumed. Still, such work remains comparatively rare relative to the volume of application papers.

This matters because the right evaluation criteria differ across tasks. In early-warning settings, one should care about warning horizon, false-positive rate, calibration, robustness to preprocessing, and performance under nonstationarity, not only whether a topological signal becomes large around a known transition \cite{George2023a,Scheffer2009a}. In neuroscience, one should care about reproducibility across participants, interpretability of state transitions, and comparison to strong dynamic-connectivity baselines, not merely whether a Mapper graph looks behaviorally plausible \cite{Saggar2018a,Saggar2022a}. In anomaly detection, the issue is not only detection accuracy, but online feasibility, stability under concept drift, and integration with existing monitoring architectures \cite{Chazal2024a,Moitra2023a}. A mature benchmarking culture would require precisely these kinds of domain-specific standards.

There is also a deeper evaluation problem. Because topology is so sensitive to representation, comparing methods fairly often requires comparing \emph{pipelines}, not isolated descriptors. A persistence image built from a poor metric can underperform a conventional feature built from a good one, while the reverse can also happen. That means topological methods should be benchmarked not only against baseline classifiers or detectors, but against alternate representations of the same underlying system. Without that, it is difficult to know whether observed gains are coming from topology itself, from a clever preprocessing choice, or from some combination of the two.

For the purposes of this review, this limitation has an important consequence. The most convincing future literature will not be literature that merely adds more applications. It will be literature that compares topological pipelines to strong competing approaches under realistic scientific constraints and reports clearly where the gains, losses, and tradeoffs lie.

\subsection{Moving from limitations to an agenda}

Taken together, these limitations define a research agenda rather than a verdict against the field. Topological methods in complex-systems science are constrained by irreducible representation dependence, incomplete inferential theory, unresolved interpretability problems, persistent computational bottlenecks, the narrowness of one-parameter workflows, and an uneven benchmarking culture. But none of these difficulties implies that topology has failed. On the contrary, they indicate that the field has progressed far enough for its foundational and methodological bottlenecks to become precise.

This is, in a sense, a sign of maturity. Early stages of a methodology are dominated by demonstrations of possibility. More mature stages are dominated by questions of validity, scope, efficiency, and explanation. Topological methods for complex systems have now clearly entered the latter stage. The next task is therefore not simply to apply topology more widely, but to develop the conceptual, statistical, computational, and evaluative machinery that allows topological results to bear more scientific weight.

The most productive way to read the present limitations is thus as prompts: how to design better null models for structurally complex data; how to connect diagrams to semantically meaningful representatives; how to scale topology to streaming and multiparameter settings; how to benchmark topology against strong alternatives; and how to integrate topological observables with mechanistic theory rather than treating them as \textit{post hoc} descriptors. The next section takes up this challenge directly by turning these open problems into a more explicit research agenda for topology in complex-systems science.

\section{Toward a Research Agenda}
\label{sec:research_agenda}

The proposed research agenda that follows is heterogeneous: some directions concern stronger invariant-based inference, others better organizational summarizers, and still others the design of higher-order relational objects and operators. Keeping those roles distinct matters because the relevant standards of success differ across them. The limitations surveyed in the previous section do not point toward retreat. They point toward a change in ambition. The first wave of topological work in complex systems was largely demonstrative: it showed that persistence, Mapper, simplicial complexes, and related constructions could reveal structure that standard summaries often miss. The next wave should be integrative. If topology is to mature into a durable part of complex-systems science, then it must move beyond retrospective case studies and become more tightly connected to representation design, dynamical modeling, causal explanation, intervention, and decision support \cite{Torres2021a,Wasserman2018a,Chazal2021a,Su2026a}.

That shift is also conceptually consistent with the central thesis of this review. If topology matters because it provides a language for emergent organization, then the field's real frontier is not simply to compute more persistence diagrams. It is to determine how topological organization should be represented, how it interacts with dynamics, how it can be manipulated or controlled, how it should be integrated with learning and inference, and in which real-world systems it is most likely to change scientific practice.

The agenda proposed here is deliberately uneven in evidential status. Some directions follow directly from limitations already visible in the established literature---better representation choice, stronger inferential procedures, tighter links to dynamics, and more rigorous comparative benchmarking. Other directions, especially topology-aware AI, streaming decision support, and large socio-technical applications, are more exploratory. They are included not as settled frontiers at which topology has already demonstrated comparative advantage, but as demanding contexts in which the field's central claims about organization, robustness, and higher-order structure can be tested most seriously. Their value is therefore partly heuristic: they identify where the method may matter, while also making clear that the present evidence base remains thinner.

\subsection{From fixed representations to adaptive and plural representations}

A first priority is to stop treating the metric, filtration, or lifting as if it were a settled preprocessing choice. Much of the practical power of topology comes from the fact that it can preserve organization across scale, but that same strength becomes a liability when the underlying representation is fixed too casually. Damrich, Berens, and Kobak show that naive Euclidean persistent homology can fail badly in high-dimensional noisy settings even when the intrinsic topology is simple, whereas better relational constructions can recover the relevant signal \cite{Damrich2024a}. Their result should be read as more than a technical warning about distance choice. It is evidence that the question ``what is the topology of the data?'' is often inseparable from the question ``what is the right representation of the system?''

For complex-systems research, this suggests a concrete agenda: topological analysis should become more \emph{representation aware} and, where possible, more \emph{representation adaptive}. In some settings, this will mean learning a metric or affinity structure before computing persistence. In others, it will mean comparing several scientifically plausible filtrations rather than defaulting to one. In still others, it will mean explicitly modeling multiple notions of scale simultaneously. Botnan and Lesnick's survey of multiparameter persistence is especially relevant here because it shows both why one-parameter filtrations are often too narrow for complex systems and why richer alternatives remain methodologically demanding \cite{Botnan2023a}. Carlsson and de Silva's zigzag persistence points in a complementary direction by accommodating systems in which structure can both appear and disappear over time, rather than only accumulate monotonically \cite{Carlsson2010a}. Together, these frameworks suggest that the next generation of topological complex-systems work should be less committed to one canonical filtration and more open to plural representations of scale, time, uncertainty, and interaction strength.

This is not a call for unconstrained flexibility. On the contrary, the field needs stronger standards for comparing alternate topological representations of the same system. A promising research program would ask, for a given scientific problem, which filtration or lifting preserves the mechanism of interest, which one is most stable under sampling variation, and which one best aligns with downstream inferential goals. In that sense, representation choice itself should become an object of empirical study rather than an unexamined prelude to analysis \cite{Torres2021a,Chazal2021a}.

\subsection{From structural description to topological--dynamical hybrids}

A second priority is to couple topology more tightly to dynamics. Much of the current literature still uses topological summaries in a largely descriptive way: a persistence diagram is computed from a reconstructed attractor, a dynamic network, or a sequence of windows, and then interpreted as evidence that the system has reorganized. That strategy has been fruitful, but it leaves a substantial gap between structure and process. If topology is truly to become a language for emergent organization, then it should increasingly participate in models of how organization evolves and how intervention changes that organization.

Several existing strands make this direction realistic. Parzanchevski and Rosenthal show that random walks on simplicial complexes connect higher-order topology to spectral and dynamical properties \cite{Parzanchevski2017a}. Schaub and colleagues extend this operational perspective by showing how normalized Hodge Laplacians support edge-space diffusion, random walks, and signal-processing constructions on simplicial complexes \cite{Schaub2020a}. Gambuzza and collaborators then demonstrate that synchronization stability depends directly on simplicial higher-order structure rather than only on a graph backbone \cite{Gambuzza2021a}. Recent control-theoretic work goes further still: Ma and colleagues derive controllability conditions for higher-order networks, while Xia and Xiang formulate pinning control directly on simplicial complexes \cite{Ma2024a,Xia2024a}. Isufi and colleagues' review of topological signal processing and learning helps tie these strands together by arguing that signals on flows, edges, and higher-order cells require an analytic language richer than node-based graph processing \cite{Isufi2025a}.

The research opportunity is to treat these results not as separate subfields, but as components of a common program in \emph{topological--dynamical hybrids}. At a minimum, three problems stand out. First, topological observables should be linked more explicitly to state variables, slow manifolds, or reduced coordinates in dynamical models, rather than being appended only after simulation or measurement. Second, higher-order operators such as Hodge Laplacians should be integrated more systematically with control, estimation, and observability theory, especially in systems whose relevant signals live on relations rather than on nodes. Third, topological summaries should increasingly be used not only to diagnose reorganization, but to ask how reorganization can be induced, delayed, or suppressed.

In other words, the field should move from asking whether a system's topology changes to asking what role topology plays in the system's reachable dynamics. That shift would substantially deepen the current literature on anomaly detection and early warning. A topological signal would no longer be only a marker of state change; it would become part of a model of which transitions are possible, which are likely, and which can be controlled.

\subsection{From structural association to causal explanation}

A third priority is to bring topology into closer contact with causal inference and causal discovery. The literature reviewed in earlier sections already shows that topological summaries can detect transitions, classify regimes, and reveal higher-order organization. What they usually do \emph{not} establish is which interactions or interventions are responsible for those topological changes. This is the point at which a mature science of topological complex systems must go beyond structural association.

The broader causal-discovery literature is already highly developed, and surveys by Zanga, Ozkirimli, and Stella make clear that many tools now exist for recovering or comparing causal structure from observational data under a range of assumptions \cite{Zanga2022a}. What remains underdeveloped is the interface between those tools and explicitly topological observables. Recent work suggests that such an interface is beginning to emerge. Smith and colleagues combine topological data analysis with causal discovery in active-matter systems to study multi-scale causality in self-organizing dynamics \cite{Smith2025a}. Kim and Lee go further by proposing explicit topological causal estimands, together with doubly robust estimation procedures, for treatment effects defined through the topological structure of outcomes rather than through changes in mean values alone \cite{Kim2026a}. These papers remain early, but they are important because they move topology from a descriptive role toward an interventional one.

For complex-systems science, at least three causal questions follow naturally. First, one can treat topological summaries as \emph{outcomes}: does an intervention change the organization of a state space, dependence structure, or higher-order interaction pattern? Second, one can treat topological structure as a \emph{mediator}: do interventions alter a system's dynamics by reorganizing its cycles, cavities, or connectivity across scale? Third, one can use topological constraints to guide \emph{causal model construction}: which causal hypotheses are plausible once the system's higher-order or multiscale organization has been taken seriously?

These questions are especially compelling for systems in which the scientific target is clearly structural. In contagion on simplicial complexes, for example, one may ask whether removing a small set of higher-order interactions changes the topological pathways through which collective adoption becomes bistable. In ecological or financial systems, one may ask whether interventions that appear small in magnitude are large in structural effect because they alter the global arrangement of the state space. In socio-technical systems, one may ask whether changes in information architecture or interdependence patterns have causal effects that are topological before they are scalar. A mature topological science of emergence should eventually be able to pose and test such counterfactuals.

\subsection{From offline analysis to streaming and decision-grade topology}

A fourth priority is operational. For topology to matter in monitoring, resilience, and decision support, it must become more naturally compatible with streaming data, concept drift, uncertainty quantification, and real-time constraints. Many current applications remain offline: windows are fixed in advance, diagrams are computed retrospectively, and interpretation occurs after the transition or anomaly is already known. That is scientifically useful, but it is not yet sufficient for deployment in infrastructures, cyber-physical systems, or other live complex environments.

There are already promising beginnings. Moitra, Malott, and Wilsey develop streaming approaches to persistent homology via topological data summaries \cite{Moitra2023a}. Chazal, Levrard, and Royer frame topological anomaly detection for dependent sequences in a way that is much closer to operational time-series monitoring than to retrospective visualization \cite{Chazal2024a}. Wang and colleagues push this integration further in industrial monitoring by using persistent homology to regularize a dynamic graph-attention model for IIoT anomaly detection, thereby turning topological structure into a constraint on learned system relationships rather than only a\textit{ post hoc} descriptor \cite{Wang2026a}. At a more formal level, work on topological signal processing shows how relation-level and flow-level signals might be processed continuously on higher-order domains \cite{Isufi2025a}.

The next step is to make these efforts more coherent. For many applications, the relevant pipeline will have the form
\[
x_{1:t}
\longrightarrow
\mathcal{R}_t
\longrightarrow
T_t
\longrightarrow
s_t
\longrightarrow
a_t ,
\]
where \(x_{1:t}\) denotes the incoming data stream, \(\mathcal{R}_t\) a time-dependent representation of the system, \(T_t\) a topological summary, \(s_t\) a warning or anomaly score, and \(a_t\) an action, decision, or recommendation. This is a much more demanding setting than classical offline TDA because each stage must now be stable, computationally feasible, and interpretable under time pressure. It also raises new statistical questions: how should uncertainty in \(T_t\) propagate to uncertainty in \(s_t\)? How should alerts be calibrated when the system undergoes concept drift? Which topological changes deserve intervention and which are benign?

These questions become especially pressing in safety-critical domains. De Marco and colleagues' review of quantitative resilience assessment in critical infrastructures shows how much demand there already is for rigorous, structural monitoring of interdependent systems \cite{DeMarco2025a}. Topology is well suited to such problems precisely because failures in these systems are often reorganizations of interdependence before they are simple point failures. But if topological methods are to be used in this setting, they must become faster, more uncertainty aware, and more tightly connected to decisions than much of the current literature is.

\subsection{From topological feature engineering to topology-aware AI}

A fifth priority is methodological synthesis with machine learning and AI. A large share of current topological machine-learning practice still treats topology as a handcrafted feature layer: compute persistence, vectorize it, and append it to an otherwise standard pipeline. That strategy has produced genuine successes, especially in low-data or structurally rich settings \cite{Turkes2022a,DiVia2024a}. But it is unlikely to be the endpoint. The more interesting research program is to build \emph{topology-aware} learning systems in which topological structure is part of the model's representational bias, inductive constraints, or objective function.

Several literatures are converging on this possibility. Zia and colleagues review the emerging field of topological deep learning and emphasize that topology can enter learning systems as input structure, regularizer, analytic lens, or architectural principle \cite{Zia2024a}. Papamarkou and coauthors go further by explicitly framing topological deep learning as a frontier for relational learning beyond ordinary graphs \cite{Papamarkou2024a}. Su and colleagues' broad review of TDA and topological deep learning beyond persistent homology reinforces the point that the field is no longer confined to \textit{post hoc} persistence features \cite{Su2026a}. Wang and colleagues' PHGAT model is a concrete example of this trend in a dynamic, real-world setting: topological summaries are not merely concatenated to learned features, but help regularize the graph that the model itself learns over time \cite{Wang2026a}.

The research agenda here is twofold. First, topology should be integrated more deeply into models that already handle sequence, graph, and higher-order data. This includes differentiable topological losses, combinatorial-complex neural architectures, topological regularizers for learned relations, and hybrid models in which topological constraints stabilize dynamic representation learning. Second, the field needs stronger comparative standards. Topology-aware AI should not be evaluated only by whether it can be made to work, but by when it improves robustness, data efficiency, interpretability, or out-of-distribution stability relative to strong non-topological baselines \cite{Turkes2022a,DiVia2024a}.

This is also the point at which the earlier themes of the review come back into focus. If topology is a language for emergent organization, then topology-aware AI is not simply AI with another feature family attached. It is an attempt to build learning systems whose internal representations are more sensitive to multiscale and higher-order organization. The challenge is to do this without sacrificing the very interpretability and structural grounding that made topology attractive in the first place.

\subsection{From familiar application areas to socio-technical grand challenges}

A sixth priority concerns domain choice. To date, many of the strongest topological applications have come from settings that are relatively tractable from a mathematical point of view: synthetic dynamical systems, curated brain-imaging datasets, historical market data, or controlled materials-science problems. Those domains have been essential for establishing proof of concept. But some of the most consequential future applications are likely to lie in messy socio-technical systems where human, informational, institutional, and infrastructural processes are tightly coupled.

Smaldino and colleagues' framework of information architectures is useful here because it makes explicit that modern societies operate on intertwined informational, social, and technical substrates whose organization shapes collective behavior and vulnerability \cite{Smaldino2025a}. De Marco and colleagues' review of resilience assessment in critical infrastructures shows, from a different angle, how urgently structural monitoring and resilience analysis are needed in interdependent systems \cite{DeMarco2025a}. These literatures are not themselves topological, but they identify a class of systems for which topological methods seem especially well matched: systems that are multiscale, relational, partially observed, and prone to failure through reorganization rather than through a single catastrophic variable excursion.

This suggests a strategically important application agenda. Topological methods should be developed more deliberately for interdependent infrastructures, information ecosystems, supply networks, cyber-physical monitoring, and other complex socio-technical environments. In such settings, one may care about whether failures propagate through higher-order dependency structures, whether information architectures become topologically brittle, whether collective coordination reorganizes before visible breakdown, or whether anomaly signals arise from subtle reconfiguration across multiple interacting layers. These are precisely the kinds of questions for which ordinary scalar monitoring can be too narrow and ordinary dyadic graphs can be too coarse.

The strategic vulnerability version of this agenda can be framed more broadly and more constructively as the study of resilience and vulnerability in systems whose failure carries societal or strategic consequence: energy, logistics, communication, public-health infrastructures, and information architectures. Topology is unlikely to solve these problems on its own. But these systems may become the most demanding and therefore the most revealing test cases for whether topological methods can scale from mathematical promise to operational science.

\subsection{Toward cumulative topological science}

Across all of these directions, one theme recurs: the field must become more cumulative. That means stronger comparative benchmarks, more realistic null models, clearer uncertainty quantification, better software for large and streaming systems, and more explicit integration with neighboring literatures in dynamics, causal inference, signal processing, and AI \cite{Bobrowski2023a,Vishwanath2025a,Turkes2022a,DiVia2024a}. The most important shift is cultural as much as technical. The field should increasingly reward studies that clarify \emph{when} topology adds value, \emph{why} it adds value, and \emph{what} kinds of organization it is actually recovering.

The agenda can therefore be summarized in two layers. A near-term methodological layer concerns adaptive representation design, stronger topological--dynamical integration, better null models and inferential procedures, clearer benchmarking, and more usable higher-order and multiparameter tools. These are direct extensions of problems already visible in current practice. A second, more exploratory layer concerns settings such as streaming monitoring, learning systems, and large socio-technical infrastructures, where topology may eventually prove especially valuable but where evaluation criteria are still being worked out. Framed in this way, the point of the agenda is not to announce a fully established program in advance of the evidence. It is to distinguish the directions that are already methodologically urgent from those that are promising precisely because they subject topological claims to harder scientific and operational tests.

If the early question for the field was whether topology could reveal previously hidden structure, the next question is more ambitious: can topology become part of the explanatory and decision-making machinery of complex-systems science itself? The answer will depend less on discovering yet another application and more on whether the field can build the theoretical, computational, and comparative infrastructure needed to make topological claims cumulative, interpretable, and actionable. That broader perspective sets up the conclusion of this review. The final step is to return to the central claim and ask what, after all of these developments, topology contributes to the scientific study of complex systems.

\section{Conclusion}
\label{sec:conclusion}

This review began from a representational problem. Complex systems are difficult to study not only because they are nonlinear, multiscale, and often nonstationary, but because the scientifically relevant structure of such systems is frequently not visible at the level of individual components, pairwise interactions, or low-order summary statistics. Simon, Anderson, and Torres and colleagues, in different ways and for different purposes, all point toward the same underlying difficulty: a system can be fully specified at one level and still remain poorly described at the level where its most important organization actually appears \cite{Simon1962a,Anderson1972a,Torres2021a}. The central claim of this review has been that topology has become increasingly useful in complex-systems science because it addresses exactly this difficulty. It provides a way to represent organization that is distributed, multiscale, and often only partially visible in conventional descriptions.

From that perspective, the significance of topology is broader than the success of any one method. Persistent homology, Mapper, simplicial complexes, clique complexes, and higher-order topological operators should not be understood as unrelated technical tools gathered under a common label. They form a family of representational strategies for asking which structures remain visible when one varies scale, relaxes metric detail, or moves beyond pairwise interaction \cite{Carlsson2009a,Edelsbrunner2008a,Wasserman2018a,Singh2007a,Giusti2016a,Bick2023a}. That shared logic is what makes the literature coherent. Topology is useful not because complex systems are secretly ``made of holes,'' but because many complex systems are organized by relations, closures, recurrences, and constraints that are not well preserved by scalar summaries or dyadic graphs alone.

This is also why the language-of-emergence formulation is more than a rhetorical flourish. Emergent organization is rarely interesting because one variable becomes large or because one interaction becomes unusually strong. It is interesting because a collective pattern becomes stable enough, and structured enough, to deserve description at its own level. Crutchfield's account of emergence as a model-building problem and Rupe and Crutchfield's more recent synthesis of emergent organization both suggest that the challenge is not only to detect novelty, but to find a mathematically explicit language in which organized novelty can be described \cite{Crutchfield1994a,Rupe2024a}. Topology contributes to this challenge by preserving connectedness, recurrence, cavity structure, and persistence across scale. In that sense, it does not replace dynamics or mechanism. It helps identify the level of organization at which a dynamical or mechanistic explanation should be sought.

The application literature reviewed here supports this interpretation. In nonlinear dynamics, topological summaries of reconstructed state spaces can detect bifurcations, transitions, and changing attractor structure \cite{Perea2015a,Mittal2017a,Myers2019a}. In finance, persistent summaries can reveal market-wide reorganization and systemic stress that are not naturally described by univariate indicators alone \cite{Gidea2018a,Ismail2022a,Ismail2022b}. In neuroscience, topological methods have been used to recover homological scaffolds, cliques, cavities, and state-transition structure in distributed neural data \cite{Petri2014a,Sizemore2018a,Saggar2018a,Saggar2022a}. In ecology, materials science, and anomaly detection, topology has likewise been effective when the scientifically relevant change is a reconfiguration of relations, not merely a local excursion in one measured quantity \cite{Larson2023a,Donato2016a,Chazal2024a}. The diversity of these applications suggests that the usefulness of topology is not tied to one special class of data, but to a more general fit between topological representations and questions about organized system-level change.

At the same time, the review has argued against a triumphalist reading of this success. Topology is not a universal replacement for statistics, graph theory, or geometry. Statistics remains indispensable when the central problem is uncertainty, estimation, or inference on scalar or distributional quantities. Graph theory remains indispensable when the natural object of interest is pairwise connectivity, transport, or community structure. Geometry remains indispensable when local metric form, curvature, or embedding is the scientific target \cite{Wasserman2018a,Newman2003a,Meila2024a}. Topology enters most naturally when the scientific burden falls on robust qualitative organization across scale or interaction order. Its proper role is therefore complementary rather than imperial. The strongest case for topology is not that other methods fail absolutely, but that they often compress away precisely the structures that some complex-systems questions most urgently require.

This more modest claim is also the more durable one because it accommodates the field's real limitations. Representation dependence remains irreducible; a persistence diagram is only as meaningful as the metric, filtration, or lifting from which it was constructed \cite{Torres2021a,Chazal2021a,Damrich2024a}. Statistical inference remains difficult and strongly dependent on scientifically appropriate null models \cite{Fasy2014a,Bobrowski2023a,Vasa2022a}. Interpretability remains incomplete because topological summaries do not automatically identify unique or semantically transparent representatives in the underlying data \cite{Li2021a,Obayashi2023a}. Computational scaling remains a practical bottleneck, especially for streaming or repeated-window analyses \cite{Otter2017a,Bauer2021a,Malott2023a}. Finally, many of the most interesting complex systems are organized by multiple parameters, non-monotone evolution, and higher-order relational structure that strain the boundaries of the classical one-parameter persistence paradigm \cite{Carlsson2010a,Botnan2023a}. These are not signs that topology has failed. They are signs that the field has reached the point where its hardest scientific questions can now be stated clearly.

For that reason, the most important conclusion of this review is forward looking. The next phase of research should not be satisfied with additional demonstrations that topology can recover structure in yet another dataset. It should ask when alternate topological representations of the same system are scientifically justified; how topological observables can be linked more tightly to dynamics, causality, and intervention; how higher-order operators and multiparameter formalisms can be made more usable; and how topological methods can be integrated into streaming, decision-oriented, and learning-based pipelines without losing interpretability or structural grounding \cite{Schaub2020a,Botnan2023a,Turkes2022a,Zia2024a,Papamarkou2024a}. If this agenda succeeds, topology will no longer appear merely as an elegant descriptive layer added after the fact. It will become part of the explanatory and operational machinery of complex-systems science itself.

The argument of this review can therefore be stated in a stronger and more defensible form. Topology is not the unique language of complex systems, and it is not a general language for emergent organization in every philosophical sense of that term. Its scientific value lies in something more specific: when the object of interest is emergent organization---when structure is multiscale, higher-order, relational, and robust under deformation---topology offers one of the most natural mathematical languages available for making that organization visible, comparable, and, increasingly, operational. On this reading, topology is not a metaphysics of emergence. It is a representational framework for studying organized system-level structure once that structure has been rendered in a form that topological methods can legitimately track.

\printbibliography

\end{document}